\documentclass{elsart}
\usepackage{graphics}
\usepackage{graphicx}
\usepackage{latexsym}
\usepackage{epsfig}
\usepackage{amssymb}
\usepackage{here}
\begin{document}

\newcommand{\bea}{\begin{eqnarray}}
\newcommand{\eea}{\end{eqnarray}}
\newcommand{\be}{\begin{equation}}
\newcommand{\ee}{\end{equation}}

\newcommand{\ed}{\end{document}}
\newcommand{\nn}{\nonumber}
\newcommand{\nnnl}{\nonumber\\}

\newcommand{\lgl}{\langle}
\newcommand{\rgl}{\rangle}
\newcommand{\del}{\partial}
\newcommand{\ccdot}{\hskip-0.3ex\cdot\hskip-0.3ex}
\newcommand{\msbar}{$\overline{\rm MS}\,\,$}
\newcommand{\msbareq}{\overline{\rm MS}}
\newcommand{\chpt}{ChPT\,\,}
\newcommand{\chpteq}{\chi PT}
\newcommand{\ve}{\varepsilon}

\renewcommand{\theequation}{\arabic{section}.\arabic{equation}}

\begin{frontmatter}

\title{Low-energy photon-photon collisions 
to two loops revisited}

\author[Bern]{J. Gasser},
\author[Dubna]{M.A. Ivanov},
\author[Helsinki]{M.E. Sainio}

\address[Bern]{Institute for Theoretical Physics, University of Bern,
Sidlerstrasse 5, \\
CH-3012 Bern, Switzerland}
\address[Dubna]{Laboratory of Theoretical Physics,
Joint Institute for Nuclear Research, \\
141980 Dubna (Moscow region), Russia}
\address[Helsinki]{Helsinki Institute of Physics, P.O. Box 64, 00014 
University of Helsinki, Finland and
Department of Physical Sciences, University of Helsinki, Finland}

\begin{abstract}
In view of ongoing experimental activities to determine the pion 
polari\-za\-bilities, we have started to recalculate 
the available two-loop expressions in the framework of 
chiral perturbation theory,  because they have 
never been checked before.
We  make use of the chiral Lagrangian at order $p^6$ now available, 
and of improved   techniques to evaluate the two-loop diagrams. 
Here, we present the result for the neutral pions.
The cross section for the reaction
 $\gamma\gamma\rightarrow\pi^0\pi^0$ agrees with the earlier calculation 
within a fraction of a percent. We present analytic results for the dipole and
quadrupole polarizabilities, and compare the latter with a recent 
evaluation {}from data on  
$\gamma\gamma\rightarrow \pi^0\pi^0$.

\vspace{0.5cm}
\end{abstract}

\end{frontmatter}

\setcounter{equation}{0}

\section{Introduction}
We consider the process $\gamma\gamma\rightarrow\pi^0\pi^0$ in the
framework of chiral perturbation theory (\chpt) \cite{GLann,GLnpb}. 
The one-loop calculation of 
the scattering  amplitude was performed
in Refs.~\cite{BijnensCornet,Donoghuepipi}, and the two-loop
amplitude  was worked out in \cite{BGS}.
Because the effective Lagrangian at order $p^6$ was not 
available at that time, the ultraviolet divergences were 
evaluated in the \msbar
 scheme,  then dropped and replaced with a corresponding polynomial
 in the external momenta. The three new counterterms which enter
at this order in the low-energy expansion were estimated
with resonance saturation. Whereas such a procedure is  legitimate
{}from a technical point of view, it does not make 
use of the full information
provided by chiral symmetry. The evaluation of the
two-loop amplitude involving charged pions was performed 
later by Burgi \cite{Burgi}. 

Over the last ten years, considerable progress has been made in this field,
 both in theory and experiment. As for theory, 
the Lagrangian at order $p^6$
 has been constructed \cite{SchererFearing,BCE1}, and its divergence 
structure has been determined \cite{BCE2}. 
This provides an important check on the above calculations: adding the
 counterterm contributions {}from the $p^6$ Lagrangian to the  \msbar 
amplitude evaluated in \cite{BGS} and in 
\cite{Burgi} must provide a scale independent result.
 Also in the  theory, improved techniques to evaluate the two-loop diagrams 
that occur in these amplitudes have been developed \cite{GS}.
The improvement arises mainly in the 
evaluation of diagrams with four external legs, 
where the techniques of Ref.~\cite{GS} allow one to 
extract the ultraviolet divergences by use of simple recursion
 relations. We are now able to present the final result 
for the two-loop amplitudes in a rather compact form 
(in Refs. \cite{BGS,Burgi}, 
the result was presented partly 
in numerical form only, because the algebraic expressions 
were too long to be published).

 Concerning experiment,  quadrupole
 polarizabilities \cite{Guiasu Radescu} 
for the neutral pions have recently been determined 
{}from data on $\gamma\gamma\rightarrow\pi^0\pi^0$ \cite{Filkovquadrupole}.
 Further, the charged pion polarizabilities 
$(\alpha-\beta)_{\pi^+}$ have  been determined  at the Mainz 
Microtron MAMI \cite{MAMI}, with a  result that 
is at variance with the two-loop calculation 
presented in \cite{Burgi}. Last but not least, there is an ongoing 
experiment by the COMPASS collaboration at CERN to measure the 
charged pion  and kaon polarizabilities \cite{COMPASS,COMPASS1}. 

In view of these developments, 
and because the two-loop expressions for the polarizabilities 
had never been checked, 
 we decided to recalculate these  amplitudes, using the 
improved techniques of Ref.~\cite{GS} to evaluate the integrals, and invoking 
the chiral Lagrangian at order $p^6$~\cite{BCE1,BCE2}.
 As the calculation in the
 case of neutral pions involves considerably less diagrams, and because the
  Fortran code for these amplitudes is still available to us for checks, 
we have decided to start the program with a re-evaluation of 
these amplitudes. This is the main purpose of the present  work.
 The evaluation of the corresponding expressions for the charged pions 
and for the kaons is underway and will be presented elsewhere 
\cite{pioncharged}.

The article is organized as follows. Section 2 contains the necessary
kinematics of the process $\gamma\gamma\rightarrow\pi^0\pi^0$.
 To make the article self contained, we summarize in Section 3 the necessary
 ingredients {}from the effective
 Lagrangian framework. In Section 4, we display the  Feynman diagrams 
and discuss their evaluation. Section 5 
contains a concise representation of the two Lorentz invariant amplitudes
that describe the scattering matrix element. 
 In Section 6, we compare the present work 
with the previous calculation~\cite{BGS}, while
 Section 7 contains explicit 
expressions for the dipole and quadrupole polarizabilities valid at 
next-to-next-to-leading order in the chiral expansion, 
together with a numerical analysis and a comparison with an evaluation {}from
data on $\gamma\gamma\rightarrow\pi^0\pi^0$~\cite{Filkovquadrupole}.
 The summary and an outlook are given in Section 8.
Finally, several technical aspects of the calculation are relegated to 
 the  Appendices.

\section{Kinematics\label{nk}}

The matrix element for the reaction
\begin{equation}
\label{eq:process}
\gamma(q_1)\, \gamma(q_2) \rightarrow \pi^0(p_1)\, \pi^0(p_2)
\end{equation}
is given by
\begin{equation}
\label{eq:matrixelement}
\langle
\pi^0(p_1) \pi^0(p_2)\,{\rm out}\,|\,\gamma(q_1)\gamma(q_2)\, {\rm in} \rangle
= i\,(2 \pi)^4 \delta^{(4)}\left(P_f-P_i\right)\,T^N \,,
\end{equation}
with
\begin{eqnarray}
T^N  & = & e^2\, \epsilon^\mu_1 \epsilon^\nu_2\, V_{\mu \nu} \,,
\nnnl
&&\nonumber\\
V_{\mu \nu} & = & i \int\! dx e^{-i(q_1x+q_2y)}
\langle\pi^0(p_1) \pi^0(p_2)\, {\rm out}\, 
|\, T j_\mu(x) j_\nu(y)\,|\, 0\rangle.
\label{eq:Tmatrix}
\end{eqnarray}
Here $j_\mu$ is the electromagnetic current, and 
$\alpha =  e^2/4 \pi \simeq 1/137$. We  consider real photons,
$q_i^2=0$, with $\epsilon_i\cdot q_i=0$. The decomposition of the correlator
$V_{\mu \nu}$ into Lorentz invariant amplitudes reads
\begin{eqnarray}
V_{\mu \nu} &=&  A(s,t,u) T_{1\, \mu \nu} + B(s,t,u) T_{2\,\mu \nu} 
               + C(s,t,u) T_{3\, \mu\nu}  + D(s,t,u) T_{4\, \mu \nu} \,,
\label{eq:amplitude}\nonumber\\
&&\nonumber\\
T_{1\, \mu \nu} &=& \frac{1}{2}\,s\, g_{\mu \nu} - q_{1\nu} q_{2\mu} \,,
\nonumber\\
&&\nonumber\\
T_{2\, \mu \nu} &=& 2\,s\,\Delta_\mu\Delta_\nu - \nu^2\,g_{\mu\nu}
                   -2\,\nu\, (q_{1\,\nu}\Delta_\mu - q_{2\,\mu}
\Delta_\nu) \,,
\nonumber\\
&&\nonumber\\
T_{3\, \mu \nu} &=& q_{1\,\mu} q_{2\,\nu} \,, 
\nonumber\\
&&\nonumber\\
T_{4\, \mu \nu}  &=&  s\,(q_{1\, \mu} \Delta_\nu - q_{2\,\nu} \Delta_\mu)
         - \nu\,(q_{1\, \mu} q_{1\, \nu} + q_{2\,\mu} q_{2\,\nu}) \,,
\nnnl
&&\nnnl
\Delta_\mu &=& (p_1 -p_2)_\mu\, ,
\end{eqnarray}
where
\bea
s &=& (q_1 + q_2)^2,\;\;\; t = (p_1 -q_1)^2, \;\;\; u = (p_2 - q_1)^2, \;\;\;
\nu = t-u \,
\eea
are the standard Mandelstam variables. The tensor $V_{\mu\nu}$  
satisfies the Ward identities
\begin{equation}
\label{eq:ward}
q^\mu_1\, V_{\mu \nu} = q^\nu_2\, V_{\mu \nu} = 0.
\end{equation}
The amplitudes $A$ and $B$  are analytic functions of the variables 
$s,t$ and $u$, symmetric under crossing $(t,u)\rightarrow (u,t)$. 
The amplitudes  $C$ and $D$ do not contribute to the process considered here,
because $\epsilon_i\cdot q_i=0$.

\begin{figure}[ht]
\begin{center} 
\epsfig{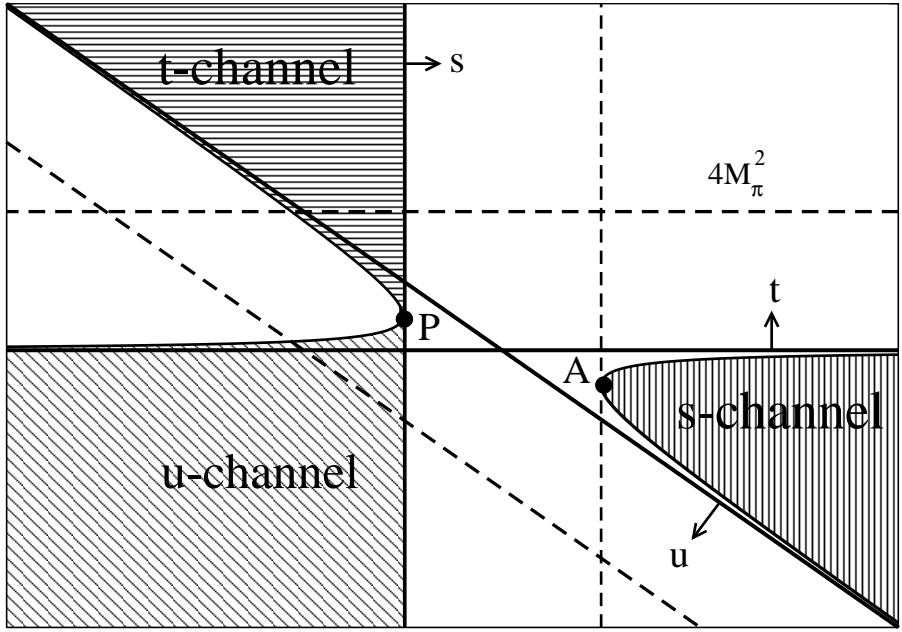} 
\caption{Mandelstam plane with  three related physical
  regions. $s$-channel: $\gamma\gamma\rightarrow \pi\pi$, $t$- and $u$-channel:
  $\gamma\pi\rightarrow\gamma\pi$. We denote the threshold for
  $\gamma\gamma\rightarrow\pi\pi\,\,\, (\gamma\pi\rightarrow\gamma\pi)$ by $A$
  ($P$). The dashed lines at $s,t,u=4\,M_\pi^2$ indicate the presence of
  branch-points in the amplitude, generated by two-pion intermediate states.}
\label{fig:kin}
\end{center}
\end{figure}

It is useful to introduce in addition the helicity amplitudes
\begin{eqnarray}
\label{eq:hel}
H_{++} &=& A + 2\,(4\,M^2_\pi - s)\, B \,,
\qquad
H_{+-} =  \frac{8\,(M_\pi^4- t\,u)}{s}\, B.
\end{eqnarray}
The helicity components $H_{++}$ and $H_{+-}$ correspond to
photon helicity differences $\lambda = 0,2$, respectively.
 With our normalization of 
states $\langle\mathbf{p_1} | \mathbf{p_2}\rangle
 = 2\,(2 \pi)^3\, p_1^0\, \delta^{(3)} (\mathbf{p_1} - \mathbf{p_2})$, the
differential cross section for unpolarized photons in the
centre-of-mass system is
\begin{eqnarray}
\label{eq:cross}
{\frac{d \sigma}{d \Omega}}^{\gamma \gamma \rightarrow \pi^0 \pi^0} 
&=& \frac{\alpha^2\,s}{64} \beta(s)\, H(s,t)\,,
\hspace{1cm}
H(s,t) = | H_{++}|^2 + | H_{+-}|^2 \,,
 \end{eqnarray}
with $\beta(s)=\sqrt{1-4\, M^2_\pi/s}$.
 The relation between the helicity 
amplitudes $M_{+\pm}$ in Ref. \cite{Filkovquadrupole} and 
the amplitudes used here is
\bea
M_{++}(s,t)=2\pi\alpha H_{++}(s,t)\,,\,M_{+-}(s,t)=16\pi\alpha B(s,t)\,.
\eea
The physical regions for the reactions $\gamma \gamma \rightarrow \pi^0 \pi^0$
and $\gamma \pi^0 \rightarrow \gamma \pi^0$
are displayed in Fig.~\ref{fig:kin},
where we also indicate with  dashed lines the nearest
singularities in the amplitudes $A$ and $B$. 
These singularities are generated by two-pion
intermediate states in the $s,t$ and $u$ channel.

\setcounter{equation}{0}
\section{The effective Lagrangian and its low energy constants} 
\label{sec:effective}
The  effective Lagrangian consists of a string of terms. 
Here, we consider QCD with two flavours, in the
isospin symmetry limit $m_u=m_d=\hat m$. At 
next-to-next-to-leading order (NNLO), one has
\bea
{\mathcal L}_{\rm eff}={\mathcal L}_2+{\mathcal L}_4 +{\mathcal L}_6\,.
\eea
The subscripts refer to the chiral order. 
The expression for ${\mathcal L}_2$ is
\bea
\label{eq:l2}
{\mathcal L}_{\, 2} &=&\frac{F^2}{4}\langle D_\mu U\,D^\mu U^\dagger
        +M^2(U + U^\dagger)\rangle\, ,\nnnl
D_\mu U &=& \partial_\mu U -i(QU-UQ)A_\mu\, , \, \
Q=\frac{e}{2}{\rm diag}(1,-1)\, ,
\eea
where $e$ is the electric charge, and $A_\mu$ denotes the electromagnetic 
field.
The quantity $F$ denotes the pion decay constant 
in the chiral limit,  and $M^2$ is the leading term in the quark mass
expansion of the pion (mass)$^2$, $M_\pi^2=M^2(1+O(\hat m))$.
Further, the brackets $\langle\ldots\rangle$ denote a trace in flavour space.
In Eq. (\ref{eq:l2}), we have retained only the terms relevant for 
the present application,
i.e., we have dropped additional external fields.
We choose the unitary $2\times 2$ matrix $U$ in the form
\bea
U &=& \sigma + i\, \pi/F\,, 
\hspace{.3cm} \sigma^2 + \frac{\pi^2}{F^2} = {\mathbf 1}_{2\times 2}\,,
\hspace{.3cm}\pi=\left( \mbox{$\begin{array}{cc} \pi^0 & \sqrt{2}\, \pi^+
         \\ \sqrt{2}\, \pi^- &- \pi^0 \end{array}$} \right)  \; .
\eea
The  Lagrangian at NLO  has the structure~\cite{GLann}
\bea
{\mathcal L}_4=\sum_{i=1}^{10} l_iK_i=\frac{l_1}{4}\langle
D_\mu U\,D^\mu U^\dagger\rangle^2
+\cdots\, ,
\eea
where $l_i$ denote low energy couplings (LECs), 
not fixed by chiral symmetry. 
At NNLO, one has \cite{BCE1,BCE2}
\bea
{\mathcal L}_6=\sum_{i=1}^{57} c_i P_i\,.
\eea
For the explicit expressions of the polynomials $P_i$, we refer the reader to
Refs.~\cite{BCE1,BCE2}. The vertices relevant for
$\gamma\gamma\rightarrow \pi^0\pi^0$ involve  $l_1, \ldots, l_6$ 
{}from ${\mathcal L}_4$ and 
$c_{29},\ldots,c_{34}$ {}from ${\mathcal L}_6$.

The couplings $l_i$ and $c_i$ absorb the divergences at order $p^4$ and 
$p^6$, respectively,
\bea\label{eq:lici}
l_i &=& (\mu\,c)^{d-4}
\left\{l_i^r(\mu,d) + \gamma_i\,\Lambda\right\}\,, 
\nn\\[2mm]
c_i&=&\frac{(\mu\,c)^{2(d-4)}}{F^2}
\left\{
c_i^r(\mu,d) - \gamma_i^{(2)}\,\Lambda^2
       -(\gamma_i^{(1)}+\gamma_i^{(L)}(\mu,d))\,\Lambda\right\}\,, 
\nn\\[2mm]
\Lambda &=& \frac{1}{16\,\pi^2 (d-4)}\,,
\, \ln c = -\frac{1}{2}\left\{\ln 4\pi +\Gamma'(1)+1\right\}\,.
\eea
The physical couplings are $l_i^r(\mu,4)$ and $c_i^r(\mu,4)$, denoted by
$l_i^r,c_i^r$ in the following.
The coefficients $\gamma_i$ are given in \cite{GLann}, 
and $\gamma_i^{(1,2,L)}$ are tabulated in \cite{BCE2}.
 In order to compare the present calculation with the result of
\cite{BGS}, we will use the scale independent quantities 
$\bar l_i$ introduced in \cite{GLann}, 
\begin{eqnarray} 
\label{RGE}
l_i^r &=& \frac{\gamma_i}{32\pi^2}\,({\bar l}_i + l)\,,
\end{eqnarray}
where the {\it chiral logarithm} is $l=\ln(M^2_\pi/\mu^2)$. We will 
use \cite{CGLpipi}
\bea
\bar l_1&=&-0.4\pm 0.6\,,\,\,\bar l_2=4.3\pm0.1\,,\,\,
\bar l_3=2.9\pm2.4\,,\,\,
\bar l_4=4.4\pm0.2\,,
\label{eq:LECsp4}
\eea
and \cite{BijnensTalavera}
\bea
\label{eq:LECs65}
\bar l_\Delta\doteq \bar l_6-\bar l_5=3.0\pm 0.3\,.
\eea
The constants $c_i^r$ occur in the combinations
 \begin{eqnarray}
a_1^r &=& 4096\pi^4\left( -c_{29}^r -c_{30}^r + c_{34}^r \right)\,,
\nn\\
&&\nn\\
a_2^r &=& 256\pi^4\left( 
8\,c_{29}^r + 8\,c_{30}^r 
            + c_{31}^r + c_{32}^r+ 2\,c_{33}^r \right)\,,
\nn \\
&&\nn\\
b^r &=&-128\pi^4
     \left(c_{31}^r + c_{32}^r + 2\,c_{33}^r \right)\,.
\label{eq:abren}
\end{eqnarray}
Their values have been estimated by resonance exchange e.g. in 
Ref.~\cite{BGS}, see also \cite{KnechtMoussallamStern}, where $c_{34}^r$
 has been determined {}from a chiral sum rule. For the present
application, we simply take the values obtained in \cite{BGS},
\bea\label{eq:LECsp6}
a_1^r(M_\rho)+8 b^r(M_\rho)&=&-14\pm 5\,,\nnnl
a_2^r(M_\rho)-2b^r(M_\rho)&=&7\pm3\, ,\nnnl
b^r(M_\rho)&=&3\pm1\, ; \, M_\rho= 770\,\, {\mbox {MeV}}.
\eea
In the numerical evaluations 
discussed later on, we use $\mu= M_\rho$. As mentioned already 
in Ref.~\cite{BGS}, varying this scale between 500 MeV and 1
GeV leads to a negligible change of e.g. the cross section
 for the reaction $\gamma\gamma\rightarrow\pi^0\pi^0$ below 
400 MeV. Finally, we will  
use $F_\pi=92.4$ MeV~\cite{Holsteinfpi} (see \cite{Descotesfpi} 
for a recent update of this value), and $M_\pi = 135$ MeV.

\setcounter{equation}{0}

\section{Evaluation of the diagrams}
\label{sec:diagrams}

\begin{figure}[!ht]
\begin{center}
\epsfig{figure=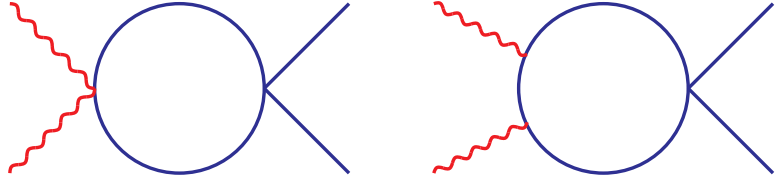,height=3cm} 
\end{center}
\caption{The one-loop diagrams.} 
\label{fig:one-loop}
\end{figure}

\begin{figure}[!ht]
\begin{center}
\epsfig{figure=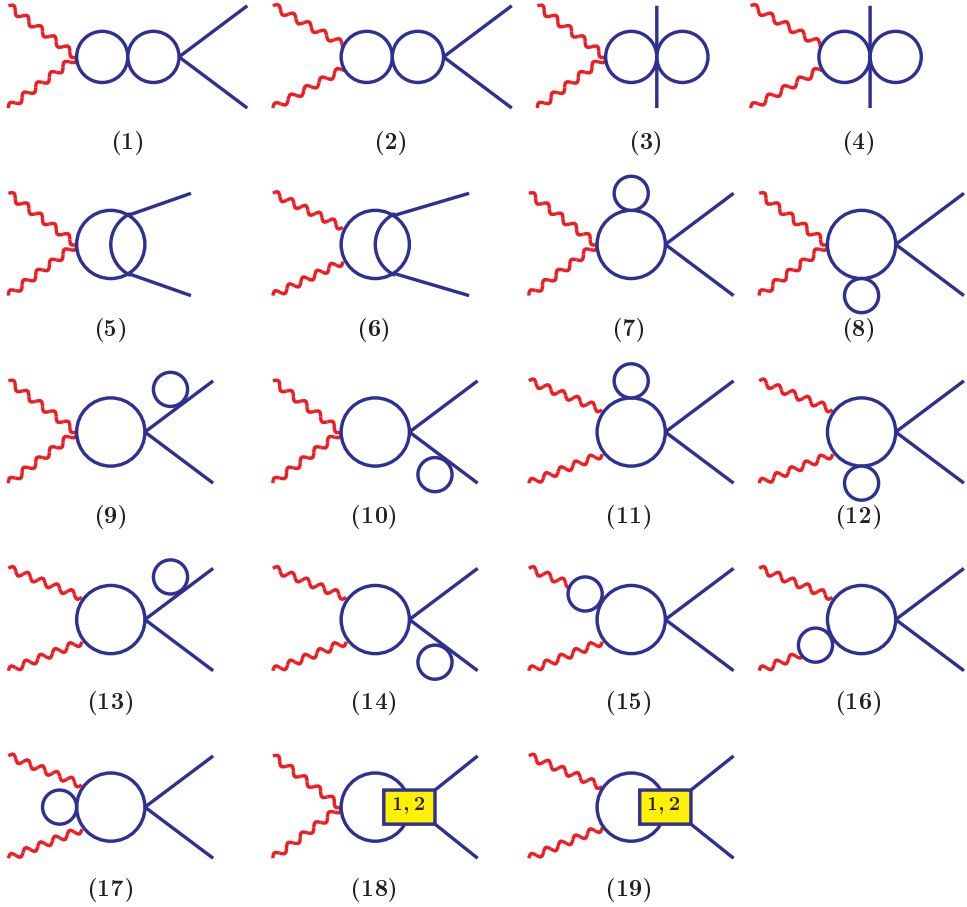,width=13.5cm}
\end{center}
\caption{A set of two-loop diagrams generated by ${\mathcal L}_2$
and one-loop diagrams generated by ${{\mathcal L}_4.}$} 
\label{fig:two-loop}
\end{figure}

\begin{figure}[!ht]
\begin{center}
\epsfig{figure=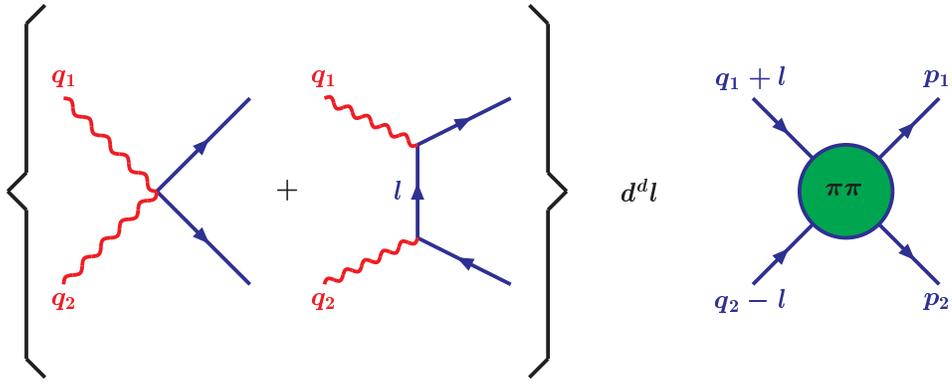,width=13.5cm} 
\end{center}
\caption{Construction scheme for the diagrams in Fig. \ref{fig:two-loop}.} 
\label{fig:scheme}
\end{figure}

\begin{figure}[!ht]
\begin{center}
\epsfig{figure=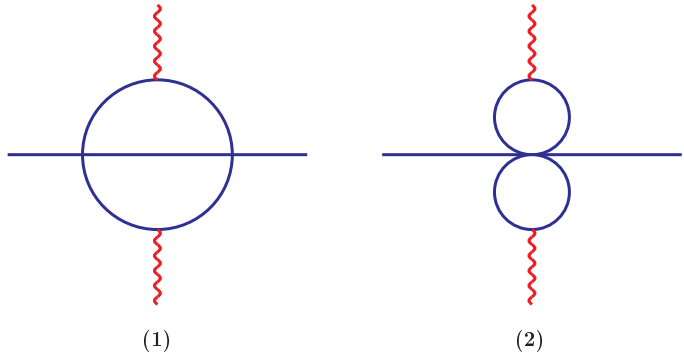,height=5cm} 
\end{center}
\caption{Acnode and butterfly diagrams.}
\label{fig:acnode}
\end{figure}

\begin{figure}[!ht]
\begin{center}
\epsfig{figure=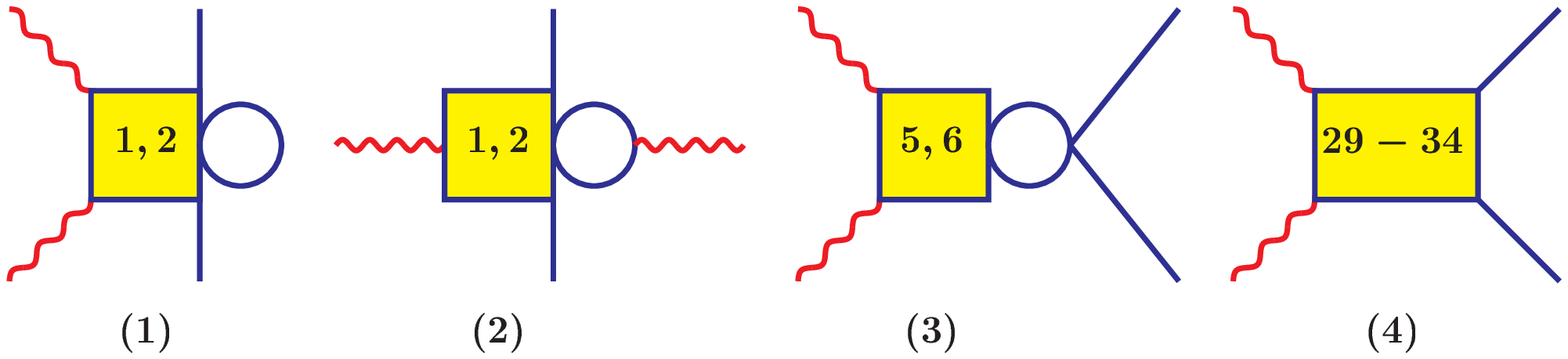,width=13.5cm}  
\end{center}
\caption{The remaining diagrams at order $p^6$: 
one-loop graphs generated by ${\mathcal L}_4$, and counterterm 
contributions  {}from ${\mathcal L}_6$.} \label{fig:lecs}
\end{figure}

The lowest-order contributions 
 are the one-loop  diagrams displayed in Fig.~\ref{fig:one-loop}. They have
 been evaluated for the first time in 
Refs.~\cite{BijnensCornet,Donoghuepipi},
 where it was noticed that the sum of these 
two amplitudes is ultraviolet
 finite, because there are no contributions {}from the effective Lagrangian at
 order $p^4$ at this order. The two-loop diagrams are displayed in the 
Figs.~\ref{fig:two-loop},\ref{fig:acnode} and \ref{fig:lecs}.
 The  two-loop
diagrams in Fig.~\ref{fig:two-loop} may be generated according to 
the scheme indicated in 
Fig.~\ref{fig:scheme}, where the shadowed blob denotes
the $d$-dimensional elastic $\pi\pi$-scattering amplitude   
at one-loop accuracy, with two pions off-shell. 
As is discussed in Appendix \ref{app:acnode}, 
the one-loop integrals in the $\pi\pi$ amplitude may be represented in a
dispersive manner. This allows one to reduce the 
two-loop integrals in Fig.~\ref{fig:two-loop} to the one-loop ones, where one
has to perform at the end an integration over a dispersive parameter.

Two further diagrams are displayed in Fig.~\ref{fig:acnode}.
The first one - called  ``acnode'' in the literature - may again be evaluated
by use of a dispersion relation, see Appendix \ref{app:acnode}.
The second one is trivial to evaluate, because it is a product of one-loop
diagrams. The remaining diagrams at order $p^6$ are shown 
in Fig.~\ref{fig:lecs}.

The evaluation of the diagrams was done in the following manner.

\begin{enumerate}
\item
We have performed the integration over the loop momenta 
in the  $d$-dimensional regularization scheme,
in particular using the procedure described 
in Ref.~\cite{GS} and in  Appendix \ref{app:acnode}, 
and invoking  FORM \cite{Vermaseren}.
\item
We have then checked numerically that the amplitude satisfies 
the Ward identities
(\ref{eq:ward}) in $d$ dimensions, in the unphysical region, 
where the amplitudes are real.
\item\label{it:calc}
We have verified that the
counterterms {}from the Lagrangian ${\mathcal L}_6$ \cite{BCE2} remove all 
ultraviolet divergences, which is a very non-trivial 
check on our calculation.
\item\label{it:calc1}
We have checked that the (ultra-violet finite) amplitude so obtained 
is scale independent.
\item
Finally, we have numerically verified that the three lowest 
partial waves of the helicity non-flip amplitude $H_{++}$ 
 carry the proper one-loop $\pi\pi$ phase, in conformity with unitarity.
\end{enumerate}
We note that the steps (\ref{it:calc}) and (\ref{it:calc1}) 
 could not  be performed in Refs. 
\cite{BGS,Burgi}, because the counterterms at order $p^6$ 
were not yet available~\cite{BCE2}.

\section{The two-loop amplitudes}
\label{sec:2-loop}
We give the expression for the amplitudes $A$ and $B$ 
by using the same notation as in \cite{BGS}, and  refer the reader to 
Appendix C of this reference for the  
one-loop integrals  $\bar{J}(s),\stackrel{=}{J}(s),
                \bar{G}(s), \stackrel{=}{G}(s)$ and  
                $\bar{H}(s)$ 
 that occur.
We have
\begin{eqnarray}
A &=& \frac{4\, \bar{G}_\pi(s)}{s\,F_\pi^2}(s-M^2_\pi)
      +U_A + P_A + O(E^4)\,,
\nonumber\\
B &=&  U_B + P_B + O(E^2)\,.
\end{eqnarray}
The quantity $\bar{G}_\pi(s)$ stands for $\bar{G}(s)$, 
evaluated with the physical pion mass.
The unitary parts $U_{A(B)}$ contain
$s,t$ and $u$-channel cuts, and $P_{A(B)}$ are  linear polynomials
in $s$. We find
\newcommand{\shift}{{}\hspace{-1cm}}
\newcommand{\shiftn}{{}\hspace{-3cm}}
\begin{eqnarray}
\shift &&
U_A = \frac{2 }{s\, F_\pi^4}\bar{G}(s) \left[ (s^2 - M_\pi^4)\,\bar{J}(s) 
     + C(s, \bar{l}_i)  \right]
     + \frac{\bar{l}_\Delta}{24\,\pi^2\, F_\pi^4}\,(s-M_\pi^2)\, \bar{J}(s)
\nonumber\\
\shift &&
  + \frac{(\bar{l}_2 - 5/6)}{144\, \pi^2\,s\,F_\pi^4}\,(s-4\,M_\pi^2)\, 
   \left \{ \bar{H}(s) + 4\,\left[s\,\bar{G}(s) +2\,M_\pi^2\,
   (\stackrel{=}{G}(s) - 3\, \stackrel{=}{J}(s))\right]\,d_{00}^2\right\}
\nonumber\\
\shift && + \Delta_A(s,t,u)\,,\nnnl
\shift &&\nonumber\\
\shift &&
C(s,\bar{l}_i) = \frac{1}{48\,\pi^2}\left\{ 
   2\, \left(\bar{l}_1 \!-\! \frac{4}{3}\right)(s \!-\! 2\, M_\pi^2)^2 
 + \frac{1}{3}\left(\bar{l}_2 \!-\! \frac{5}{6}\right)
   (4\,s^2-8\,s\,M_\pi^2+16\,M_\pi^4) \right.
\nonumber\\[2mm]
\shift && \left.
 - 3\, M_\pi^4\, \bar{l}_3 + 12\, M_\pi^2\, (s - M_\pi^2)\,\bar{l}_4 
 - 12\,s\, M_\pi^2 + 15\,M_\pi^4 \right\}\,, 
\nonumber\\
\shift &&\nonumber\\
\shift &&
d_{00}^2 = \frac{1}{2}(3\,\cos^2\theta - 1 )  \,,
\label{UA}\\
\shift &&\nonumber\\
\shift &&\nonumber\\
\shift &&
U_B = \frac{(\bar{l}_2 - 5/6)\,\bar{H}(s)}{288\, \pi^2\, F_\pi^4\, s}
         + \Delta_B(s,t,u)\,.
\label{UB}
\end{eqnarray}
The expressions for  $\Delta_{A(B)}$ 
 are displayed  in the Appendices
\ref{app:deltaab} and \ref{app:polynomials}.

The polynomial  parts are 
\begin{eqnarray}\label{eq:polynoma}
P_A &=& \frac{1}{(16\, \pi^2\, F_\pi^2)^2}\, 
        \left[a_1\, M_\pi^2 + a_2\, s \right]\,,
\nnnl
a_1 &=& a_1^r + \frac{1}{18} \left\{ 4\, l^2 + l\,\left(8\, \bar{l}_2 
     + 12\,\bar{l}_\Delta - \frac{4}{3}\right) -\frac{20}{3}\,\bar{l}_2
     + 12\,\bar{l}_\Delta  +\frac{110}{9}\right\}\,,
\nonumber\\
a_2&=& a_2^r -\frac{1}{18}\left\{l^2 +l\,\left(2\,\bar{l}_2 
     +12\,\bar{l}_\Delta -\frac{4}{3}\right)-\frac{5}{3}\,\bar{l}_2 
     +12\,\bar{l}_\Delta +\frac{697}{144} \right\} \,,
\label{PA}\\
\shift &&\nonumber\\
P_B &=& \frac{b}{(16\,\pi^2\,F_\pi^2)^2}\,,
\nnnl
b &=& b^r - \frac{1}{36} \left[ l^2 + l\,\left( 2\, \bar{l}_2 
          + \frac{2}{3}\right) -\frac{1}{3}\,\bar{l}_2 
          + \frac{393}{144} \right]\, ,\nnnl
l&=&\log{\frac{M_\pi^2}{\mu^2}}\, .
\label{PB}
\end{eqnarray}
The constants $a_1^r, a_2^r$ and $b^r$ are displayed in terms of the LECs at
order $p^6$ in Eq. (\ref{eq:abren}). Using the fact that the bare
couplings $c_i$ displayed in Eq. (\ref{eq:lici}) are scale independent, one
indeed finds that the above expressions for the amplitudes $A,B$ are scale
independent as well. 

\section{Comparison with the previous calculation}

We can now compare the amplitudes $A,B$ 
 with the earlier calculation,
presented in Section 7 of Ref.~\cite{BGS}. In that reference, the 
amplitudes were evaluated  with a
different techniques. Furthermore,  the Lagrangian ${\mathcal L}_6$
was not available in those days, and an important 
ingredient to check the final result was, therefore,  missing. 
We can make the following observations.
\begin{enumerate}
\item
The amplitudes $A$ and $B$ consist of a part with explicit 
analytic expressions, and additional terms $\Delta_{A,B}$, 
that are given in the Appendices
\ref{app:deltaab} and \ref{app:polynomials} of the present work in the 
form of integrals over Feynman parameters.
 These latter 
terms were given only in numerical form in \cite{BGS}.
\item
The explicit analytic
expressions agree with the previous calculation, except for 
the coefficient of the single logarithm in  $a_2$ in
Eq.~(\ref{eq:polynoma}). The factor 2/3 in \cite{BGS} is replaced by - 4/3 
here. As the present amplitude is scale 
independent, we conclude that it is  the result Eq. (\ref{eq:polynoma})
 which is correct\footnote{Burgi~\cite{Burgithesis}  
provides in his thesis work the isospin $I=0,2$ amplitudes, 
and the one for   the charged pions. 
Subtracting the latter {}from the former reveals that his
   calculation agrees with the statement just made.}. 
This mistake does not
 affect the algebraic expressions for the  polarizabilities
 discussed below, for which we fully agree with Ref.~\cite{BGS}.
\item
We can compare the quantities $\Delta_{A,B}$ in numerical form only. 
For this purpose, we have made two checks. First, we 
 have evaluated the cross section  for the reaction 
 $\gamma\gamma\rightarrow\pi^0\pi^0$ below a centre-of-mass
energy of 400 MeV, using the same values for the LECs as in \cite{BGS}.
 It agrees with the previous one within a fraction of a percent - the
 difference would not be  visible  in Fig. 5 of
 Ref.~\cite{BGS}, and we do not, therefore,  reproduce that plot here.
Second, we have re-evaluated the two-loop contributions 
to the polarizabilities presented in column 4 of Table 3 in \cite{BGS}.
The numbers (0.17, - 0.31) in the old calculation become (0.17, - 0.30) 
here.
\item
To summarize, we confirm the previous result up to the coefficient 
in one of the chiral logarithms, and up to minute changes in the numerical 
values of $\Delta_{A,B}$.  Numerically, the results 
in \cite{BGS} are not affected in any significant manner by 
these modifications, whose effect is by far smaller than  
the uncertainties generated by the
(not precisely known)  values of the low energy constants.
\end{enumerate}

\setcounter{equation}{0}
\section{Pion polarizabilities: dipole and quadrupole}
\label{sec:polar}
The {\it dipole} and {\it quadrupole} 
polarizabilities 
are defined \cite{Guiasu Radescu,Filkovquadrupole}
through the expansion of the helicity
amplitudes at fixed $t=M_\pi^2$,
\bea\label{eq:defpolarizabilities}
\frac{\alpha}{M_\pi}H_{+\mp}(s,t=M_\pi^2)=(\alpha_1\pm\beta_1)_{\pi^0}
+\frac{s}{12}(\alpha_2\pm\beta_2)_{\pi^0}+{\mathcal O}(s^2)\, .
\eea
Because we have at our disposal the helicity amplitudes at two-loop order,
 we can work out the polarizabilities to the same accuracy. 
It turns out that all relevant integrals can be performed in closed
form. We discuss the results in the remaining part of this Section.

\subsection{Chiral expansion}
Using the same notation as in \cite{BGS}, we find for the 
{\it dipole} polarizabilities
\begin{equation}
\label{alphabeta}
(\alpha_1 \pm \beta_1)_{\pi^0} = 
           \frac{\alpha}{16\,\pi^2\,F_\pi^2\,M_\pi}\,
\left\{c_{1\pm} + \frac{M_\pi^2\,d_{1\pm}}{16\,\pi^2\,F_\pi^2} +O(M^4_\pi)
\right\}\,,
\end{equation}
with 
\begin{eqnarray}
c_{1+} &=& 0\,,c_{1-}=-1/3\, ,\nnnl
d_{1+} &=& 8\,b^r 
     - \frac{1}{648}\,(144\,l\,(l+2\,\bar l_2)
                  +96\,l+288\,\bar l_2+ 113  + \Delta_+  )\,,
\nnnl
&&\nnnl
d_{1-} &=& a_1^r+8\,b^r
     +\frac{1}{648}\,\left(144\,l\,(3\,\bar l_\Delta-1)
                    +36\,(8\,\bar l_1-3\,\bar l_3-12\,\bar l_4
                   +12\,\bar l_\Delta)\right. \nnnl
&&\left. +43 +\Delta_-\right) \,,
\nnnl
\Delta_+ &=& 13643-1395\,\pi^2\,,\Delta_- = - 3559 + 351\,\pi^2\,.
\end{eqnarray} 
We have split off the numbers 113 and 43, respectively, to illustrate that
these expressions completely agree with the ones displayed in Eq. (8.14) of
Ref. \cite{BGS}, where, as already mentioned, no explicit expressions for the
remainders $\Delta_{+,-}$ were worked out.
For the {\it quadrupole} polarizabilities, we obtain
\begin{equation}
\label{alphabeta2}
(\alpha_2 \pm \beta_2)_{\pi^0} = 
           \frac{\alpha}{16\,\pi^2\,F_\pi^2\,M_\pi^3}\,
\left\{c_{2\pm}
 + \frac{M_\pi^2 d_{2\pm} }{16\,\pi^2\,F_\pi^2} +O(M^4_\pi)
\right\}\,,
\end{equation}
with
\bea
 c_{2+}&=&0\,,\, c_{2-}=156/45\, ,\nnnl
d_{2+}&=&
 - \frac{5009}{27} + \frac{13453\, \pi^2}{720} 
          + \frac{16\,\bar{l_2}}{45}\,,\\[2mm]
d_{2-}&=& 
12\,a_2^r - 24\,b^r
+\frac{1}{960}\left(
1280\,l\,(1-6\, \bar{l}_\Delta)
+19216 - 1811 \,\pi^2\right)\nonumber\\[2mm]
&&-\frac{4\,(52\, \bar{l}_1 + 5\,\bar{l}_2 + 3\, \bar{l}_3 
           - 78\, \bar{l}_4 + 105\, \bar{l}_\Delta)}{45}\,.
\eea
\subsection{Numerical results}

\begin{figure*}
\begin{center} 
\epsfig{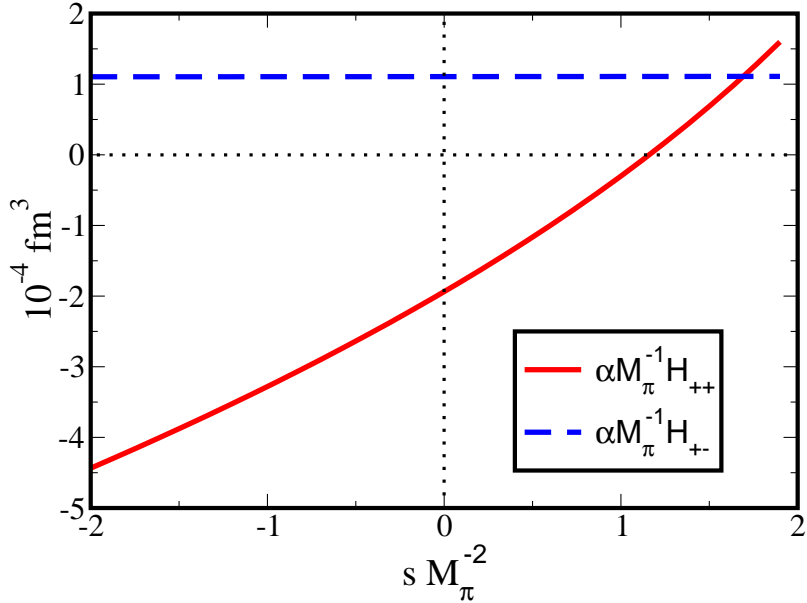} 
\caption{ The helicity amplitudes
  $\frac{\alpha}{M_\pi}H_{++}(s,t=M_\pi^2)$ (solid line),
$\frac{\alpha}{M_\pi}H_{+-}(s,t=M_\pi^2)$ (dashed line),
 plotted as a function of $s$, at  $t=M_\pi^2$. 
 Compare with the definition of
 the polarizabilities in Eq.~(\protect{\ref{eq:defpolarizabilities}}) The
 dotted lines are displayed to guide the eye.
        }
\label{fig:polar}
\end{center}
\end{figure*}

For numerical evaluations of the polarizabilities 
we use the values of the LECs given in Section
\ref{sec:effective}.
\begin{table}
\begin{center}
\caption{The dipole and quadrupole polarizabilities in units
of $10^{-4}\,{\rm fm}^3$ and $10^{-4}\,{\rm fm}^5$, respectively. 
The symbol $[\ast]$ refers to  the present work. 
The slight difference in the value of  $(\alpha_1+\beta_1)_{\pi^0}$ 
 with the one reported in \cite{BGS} is due to the updated  values of
 $\bar{l}_2$ and of the pion decay constant $F_\pi$ used here.} 
\label{tab:pol}
\def\arraystretch{2}
\begin{tabular}{|r|r|r|}
\hline
                    & \chpt         & dispersion relations  \\
\hline
$ (\alpha_1-\beta_1)_{\pi^0} $ & 
                                $ -1.9\pm 0.2$\, \cite{BGS}
                               &$ - 1.6 \pm  2.2 $\,\,\cite{Filkovdipole}
\\
\hline
$ (\alpha_1+\beta_1)_{\pi^0} $ & 
                                $  1.1 \pm 0.3$\, \cite{BGS}
                               &$  0.98 \pm 0.03 $\, \cite{Filkovdipole}
\\
                               & & $ 1.00 \pm 0.05 $\, \cite{Kaloshindipole} 
\\
\hline
$ (\alpha_2-\beta_2)_{\pi^0} $ & 
                                $ 37.6  \pm 3.3$\, $[\ast]$
                               &$ 39.70 \pm 0.02$\,\,\cite{Filkovquadrupole}
\\
\hline
$ (\alpha_2+\beta_2)_{\pi^0} $ & 
                                $  0.04$  $[\ast]$
                               &$ -0.181 \pm 0.004 $\,\,\cite{Filkovquadrupole}
\\
\hline\hline
\end{tabular}
\end{center}
\end{table}
The results are displayed 
in Table \ref{tab:pol}, where we also quote the 
results {}from dispersive calculations. The following comments are in order.
\begin{enumerate}
\item
The slight difference in the value of  $(\alpha_1+\beta_1)_{\pi^0}$ 
 with the one reported in \cite{BGS} is due to the updated  values of
 $\bar{l}_2$ and of the pion decay constant $F_\pi$ used here.
\item
Our results for the dipole
polarizabilities as well as for the quadrupole polarizabilities 
  $ (\alpha_2-\beta_2)_{\pi^0} $
 agree with the results of the recent
investigations performed in \cite{Kaloshindipole,Filkovdipole}
and \cite{Filkovquadrupole} within the uncertainties quoted. 
\item
The prediction for the
quadrupole polarizability 
$ (\alpha_2+\beta_2)_{\pi^0} $ is positive, in contrast to the result
reported in Ref.~\cite{Filkovquadrupole}. The \chpt expression 
contains as the only LEC  $\bar{l}_2$,
known rather accurately {}from $\pi\pi$ scattering \cite{CGLpipi}.
We come back to this point in 
the following subsection, where we also discuss the uncertainties 
 quoted in the Table for the \chpt calculation.
\item
We plot the helicity amplitudes in Fig.~\ref{fig:polar}.
It illustrates  the fact that the helicity flip amplitude  $H_{+-}$
is quite flat at this order, in contrast to the non-flip amplitude $H_{++}$, 
see the values of the quadrupole polarizabilities in Table \ref{tab:pol}.
\item
For a comparison of the \chpt - predictions
of  the dipole polarizabilities with calculations
performed before 1994, and for additional information 
on these quantities,   we refer the interested reader to
 Refs.~\cite{BGS} and \cite{Bellucci}.

\end{enumerate}

\subsection{Estimating the uncertainties}
To estimate the uncertainties in the prediction of the polarizabilities,
we first note that the helicity non-flip amplitude $H_{++}$ starts out at
order $p^4$. We have therefore, for this quantity, a leading and
next-to-leading order calculation at our 
disposal.
 For the corresponding
polarizabilities $(\alpha_1-\beta_1)_{\pi^0}$ and 
$(\alpha_2-\beta_2)_{\pi^0}$,   we thus
  simply add in quadrature the uncertainties generated by the
  order $p^4$ and $p^6$ LECs (see Section \ref{sec:effective}). The resulting
  numbers are given in column 2 of Table \ref{tab:pol}. They do not
  incorporate an estimate of the  higher order contributions.

On the other hand, the helicity flip amplitude $H_{+-}$ 
starts out at order $p^6$, and we have determined here only its 
leading order term. According to Eq.~(\ref{eq:hel}), this amplitude
is proportional to $B(s,t,u)$, which
is an analytic function of the variables $s,\nu$
 at the Compton threshold and can be, therefore,  expanded 
in a Taylor series,
\bea
B(s,t,u)=U+Vs+W\nu^2+{\mathcal O}(s^2,\nu^4,s\nu^2)\,.
\eea
The relation to the polarizabilities is
\bea
(\alpha_1+\beta_1)_{\pi^0}=8\alpha M_\pi U\,,\,
(\alpha_2+\beta_2)_{\pi^0}=96\alpha M_\pi V\,.
\eea
The Taylor coefficients themselves have a chiral expansion of the form
\bea
U&=&\frac{1}{(16\pi^2F_\pi^2)^2}\left[U_0+\frac{M_\pi^2U_1}
{16\pi^2F_\pi^2}+{\mathcal O}(M_\pi^4)\right]\,,\nnnl
V&=&\frac{1}{(16\pi^2F_\pi^2)^2M_\pi^2}
\left[V_0+\frac{M_\pi^2V_1}{16\pi^2F_\pi^2}
+{\mathcal O}(M_\pi^4)\right]\,.
\eea
Whereas LECs {}from order $p^6$ do contribute to $U_0$, the 
leading term $V_0$ is a pure loop effect, because $V_0/M_\pi^2$ is not
 analytic in the pion mass and thus cannot receive contributions 
{}from polynomial  counterterms.
To illustrate this point, and to estimate the size of $V_1$, 
we consider the vector meson exchange amplitudes 
worked out in \cite{BGS}.  The contribution {}from
$\omega$ exchange is dominant and given by
\bea\label{eq:ampbomega}
B_\omega(s,t,u)=\frac{C_\omega}{2}\left[\frac{1}{M_\omega^2-t}+\frac{1}{M_\omega^2-u}\right]\,;\,
C_\omega = 0.67\, {\rm GeV}^{-2}\, .
\eea
In the language of \chpt, this amplitude starts out at order $p^6$. 
It does contribute 
to $U_0$ - this term is included in the resonance exchange estimates
for the ${\mathcal O}(p^6)$ LECs in (\ref{eq:LECsp6}).
For this reason,  we calculate the
uncertainties in $(\alpha_1+\beta_1)_{\pi^0}$ as before, with a result
that is given in the second row of Table \ref{tab:pol}. Again, it doe not
incorporate an estimate of higher order contributions.

Finally, we come to the estimate of the uncertainty in 
 $(\alpha_2+\beta_2)_{\pi^0}$. In agreement with what is said above,
 resonance exchange does not contribute to $V_0$. On the other hand, there
 is no reason why this term  should dominate 
the contribution {}from $V_1$. Indeed, if we use (\ref{eq:ampbomega}) 
to also estimate effects {}from order $p^8$, we find
 with $M_\omega=782$ MeV the value $V_1=-2.2$. 
The corresponding contribution to the 
quadrupole polarizability $(\alpha_2+\beta_2)_{\pi^0}$ 
is $-0.25\cdot 10^{-4} \,{\rm fm}^5$ -  of the order needed 
to bring the \chpt calculation into agreement with the analysis
of Ref.~\cite{Filkovquadrupole}. 

This result  illustrates that the discrepancy between the
chiral prediction for  the quadrupole polarizability 
$(\alpha_2+\beta_2)_{\pi^0}$ at order $p^6$ and the dispersion 
analysis  in Ref.~\cite{Filkovquadrupole} is of no
significance, because the terms neglected may well be much larger
 than the leading order term, which would only dominate
 for very small values of the pion mass.
On the other hand,  to obtain a reliable 
estimate of $(\alpha_2+\beta_2)_{\pi^0}$ in the framework of \chpt, 
one needs to perform a reliable calculation of the relevant
couplings  at order $p^8$. This is outside the scope of 
the present work.
For this reason, we do not quote an uncertainty for
 $(\alpha_2+\beta_2)_{\pi^0}$ in Table \ref{tab:pol}.

\section{Summary and outlook}

\begin{enumerate}
\item
We have recalculated the two-loop expression for the
amplitude $\gamma\gamma\rightarrow\pi^0\pi^0$ in the framework of chiral
perturbation theory. We have made use of
 the   techniques developed in Ref.~\cite{GS}, and of the
 effective Lagrangian ${\mathcal L}_6$ available now \cite{BCE1,BCE2}.
\item
The method has allowed us to evaluate the dipole
and quadrupole polarizabilities in closed form.
[As far as we are aware, the quadrupole polarizabilities 
have never been calculated in \chpt before.]
The two Lorentz invariant amplitudes $A$ and $B$  are
 presented as a sum over multiple integrals over Feynman
parameters whose numerical evaluation poses no difficulty.
This is in contrast to Ref.~\cite{BGS}, where part of the
amplitudes, denoted by $\Delta_{A,B}$,  were published in 
numerical form only.
\item
Our result agrees with the earlier calculation \cite{BGS} up the the
coefficient in one of the chiral logarithms in the amplitude $A$, and 
up to minute differences in the numerical values of the 
remainder $\Delta_{A,B}$. The induced changes in the numerics 
of the cross section and of the dipole polarizabilities are far below 
the uncertainties generated by the (not precisely known) 
values of the low energy constants.
\item
The values for the dipole and 
quadrupole polarizabilities are presented in Table 
\ref{tab:pol} and confronted
with recent evaluations {}from data on $\gamma\gamma\rightarrow\pi^0\pi^0$.
There is reasonable agreement for the dipole
polarizabilities. As for the
quadrupole ones, the combination  $(\alpha_2-\beta_2)_{\pi^0}$ related 
to the helicity non-flip amplitude agrees with
\cite{Filkovquadrupole} within the uncertainties quoted.
On the other hand, the  sum $(\alpha_2+\beta_2)_{\pi^0}$ -
related to the helicity flip amplitude - differs in sign {}from the one in
 Ref.~\cite{Filkovquadrupole}.
 We have shown why this does not contradict the
predictions of \chpt: this quantity is a two-loop effect, and one expects {}from
order $p^8$ (three loops) substantial corrections to the leading order 
result. We have indeed identified $\omega$-exchange as an important  
contribution at this order.
\item
It would be instructive to improve the estimates for the LECs
$c_{29},\ldots, c_{34}$ in the sense
that in these estimates, the constraints {}from the asymptotics of QCD
\cite{LECs_asymptotic,KnechtMoussallamStern} should be respected. 
\end{enumerate}
The corresponding calculation of the charged pion
polarizabilities is in progress \cite{pioncharged}.

\ack
 This work was completed while M.A.I. visited
the University of Bern. 
It is a pleasure to thank G.~Colangelo, J.~Bijnens and 
B.~Moussallam for useful discussions,
and S.~Bellucci for useful remarks concerning the manuscript.
This work was  supported  by the Swiss
National Science Foundation, by RTN, BBW-Contract No. 01.0357,
and EC-Contract  HPRN--CT2002--00311 (EURIDICE).
M.A.I. also appreciates the partial support by
the Russian Fund of Basic Research under Grant No. 04-02-17370.  
Also, partial support by the Academy of Finland, grant 54038,
is acknowledged.

\appendix

\section{Notation}\label{app:notation}
In order to simplify the expressions, we set the pion mass equal 
to one in all Appendices,
\bea
M_\pi=1\, .
\eea
We use the following notation for $d$-dimensional  one-loop and 
two-loop integrals,
\bea
\langle...\rangle &=&\int\frac{d^d l}{(2\,\pi)^d\,i}\,(...)\,,
\hspace{1cm}
\langle\langle...\rangle\rangle =
\int\frac{d^d l_1}{(2\,\pi)^d\,i}\,\int\frac{d^d l_2}
{(2\,\pi)^d\,i}\,(...)\,.
\eea
In particular, 
\bea\label{eq:F_n}
\left\langle\frac{1}{[z-l^2]^n}\right\rangle&=&F_n[z]\, , n\ge 1\, ,\nnnl
F_n[z]&=& z^{w+2-n}\,C(w)\,\frac{\Gamma(n-2-w)}{\Gamma(n)}\,,\,
w = \frac{d}{2}-2\,.
\eea
The measures in the integration over Feynman parameters are defined
by 
\bea
d^2x=dx_2dx_3\,\, , \, \, d^3x=dx_1dx_2dx_3\,\,.
\eea
In dispersion relations, we use the $d$-dimensional measure
\bea
{[d\sigma]} 
&=& \frac{C(w)\,\Gamma(3/2)}{\Gamma(3/2 + w)}
 \left(\frac{\sigma}{4}-1\right)^w\,\beta\;d\sigma\,,\hspace{0.5cm} 
\nonumber\\
&&\nonumber\\
C(w)&=&\frac{1}{(4\pi)^{2+w}}\,,\hspace{0.5cm} 
\beta=\sqrt{1-4/\sigma}\,.\hspace{0.5cm} 
\label{notationGS}
\eea
 
\section{The acnode}\label{app:acnode}
The evaluation of the two-loop vertex and box diagrams (5) and (6) in 
Fig.~\ref{fig:two-loop}
 is described in Ref.~\cite{GS}.
It is based on a $d$-dimensional dispersive representation
of the fish-type diagram,
\begin{eqnarray}
J(p^2) &=& \left\langle 
\frac{1}{\left[1-l^2\right]\left[1-(l+p)^2\right]}\right\rangle
= C(w)\,\Gamma(-w)\int\limits_0^1\! dx\, [1-p^2\, x\,(1-x)]^w
\nonumber\\
&&\nonumber\\
&=& \int\limits_4^\infty \frac{[d\sigma]}{\sigma-p^2}\,\,;\,-1.5 < \omega
< 0\, ,
\label{disp1}
\end{eqnarray}
where the measure $[d\sigma]$ is given in Appendix \ref{app:notation}. 
This representation allows one to reduce the two-loop
 vertex and box integrals  to the one-loop
 case, with a final  integration
over the dispersion parameter $\sigma$. The ultraviolet
divergences can be extracted by invoking  recursion
relations~\cite{GS}. 
 While  the acnode was treated in a different manner in~\cite{GS},
we evaluate  it here  analogously to 
the vertex and box diagrams just mentioned. This results in considerable
 simplifications in the numerical programs. For this purpose, we invoke
 a dispersion relation for the function $I(m,n;s)$ defined by
\begin{eqnarray}
I(m,n;s) &=& 
\int\limits_0^1 dx \left[1-s\,x\,(1-x)\right]^m\left[x\,(1-x)\right]^n\,,
\hspace*{1cm} -1<m<0 \,.\nnnl
\label{Imn}
\end{eqnarray}
$I(m,n;s)$ is analytic in the complex $s$-plane, cut along the
real axis for Re $s\geq 4$. To evaluate its absorptive part, 
we observe that the imaginary part of
the first factor of the integrand in Eq.~(\ref{Imn}), 
evaluated at the upper rim of the cut, is 
\begin{eqnarray}
&&
{\rm Im}\,[1-s\,x\,(1-x)]^m
=-\,\sin(\pi\,m)\,\left[s\,x\,(1-x)-1\right]^m\,, 
\nnnl
&&
s>4\,\hspace*{0.5cm} x_-< x < x_+\,,
\hspace*{0.5cm}
x_\pm = (1/2)\,\left(1\pm\sqrt{1-4/s}\right)\,.
\label{ImImn}
\end{eqnarray}
The integrand is  symmetric around $x=1/2$, so we
may restrict the integration {}from $x_-$ to $1/2$ in the evaluation of the
 absorptive  part of $I(m,n;s)$. The substitution
$x=(1/2)\,\left(1-\sqrt{(1-4/s)\,(1-u)}\right)$
 generates  a hypergeometric function, and 
we arrive at the dispersion relation 
\begin{eqnarray}
I(m,n;s) &=& 
\int\limits_4^\infty \frac{d\sigma\,\rho\,(m,n;\sigma)}{\sigma-s}\,,
\nnnl
&&\nonumber\\
\rho(m,n;\sigma) &=& \frac{\Gamma(3/2)\,\delta^{1/2+m}}
                          {4^n\,\Gamma(3/2+m)\Gamma(-m)}
\,\,_2F_1\left(\frac{1}{2},\frac{3}{2}+m+n;\frac{3}{2}+m;-\delta\right)\,,
\nonumber\\
&&\nonumber\\
\delta &=& \frac{\sigma}{4}-1\,.
\label{disp2}
\end{eqnarray}
In particular, we find
\begin{eqnarray}
\rho(m,0;\sigma) &=& \frac{\Gamma(3/2)\,\beta\,\delta^{m}}
                          {\Gamma(3/2+m)\Gamma(-m)}\,,
\nnnl
&&\nnnl
\rho(m,1;\sigma) &=& \frac{\Gamma(3/2)\,\beta\,\delta^{m}}
                          {4\,\Gamma(5/2+m)\Gamma(-m)}\,
                          \left(1+m+\frac{2}{\sigma}\right)\,,
\nnnl
&&\nnnl
\rho(m,2;\sigma) &=& \frac{\Gamma(3/2)\,\beta\,\delta^{m}}
                          {16\,\Gamma(7/2+m)\Gamma(-m)}\,
 \left(2+3\,m+m^2+\frac{4\,(1+m)}{\sigma}+\frac{12}{\sigma^2}\right)\,,
\nnnl\end{eqnarray}
with $\beta$ given by Eq.~(\ref{notationGS}).

The dispersion relation (\ref{disp2})
allows us to evaluate the integral that occurs in the evaluation of the acnode
diagram (1) in Fig.~\ref{fig:acnode},

\begin{eqnarray}
A_N^{\mu\nu} &=& \langle\langle
\frac{ 4\,l_1^\mu l_2^\nu\,V_L\,V_R}
     {D_1\,D_2\,D_3\,D_4\,D_5}\rangle\rangle\,,
\nnnl
&&\nnnl
V_L &=& (l_2+q_2-l_1)^2 -1 \,,
\nonumber\\
V_R &=& (l_1+q_1-l_2)^2 -1 \,,
\nonumber\\
D_1 &=& 1-l_1^2\,,\hspace{1cm}
D_2 = 1-(l_1+q_1)^2\,,
\nonumber\\
D_3 &=& 1-l_2^2\,,\hspace{1cm}
D_4  = 1-(l_2+q_2)^2\,,
\nonumber\\
D_5 &=& 1-(l_1-l_2-p_1+q_1)^2\,.
\label{acnode}
\end{eqnarray}

We combine $1/(D_1\,D_2)$ and  $1/(D_3\,D_4)$ by using 
$x_1$ and $x_2$ as Feynman parameters, respectively.
By shifting $l_1$ and $l_2$ and again dropping terms that vanish upon
contraction with the polarization vectors, one obtains
\bea
A_N^{\mu\nu} &=& 
\int\limits_0^1\!d^2 x\, 
\langle\langle\frac{l_2^\nu}{[1-l_2^2]^2}
\cdot
\frac{l_1^\mu\,P_N(l_i,p_i,q_i)}{[1-l_1^2]^2\,[1-(l_1-r)^2]}
\rangle\rangle\,,
\nnnl
&&\nnnl
r &=& l_2-q, \hspace{1cm} q=x_1 q_1+x_2 q_2-p_1\,.\label{eq:acnode}
\eea
Here, $P_N(l_i,p_i,q_i)$ is a polynomial in the momenta indicated.
The integration over $l_1$ is performed by using
the  dispersion relation (\ref{disp2}) 
with $s=r^2$. 
Then we proceed in a manner which is similar to the case of the 
box diagram described in Ref.~\cite{GS}. 
The final expression can be  written as a combination
of the integrals
\begin{eqnarray}
A(i,k,m,n) &=& 
\int\limits_4^\infty [d\sigma]\,\int\limits_0^1\! d^3x
\left( \frac{\sigma}{4}-1 \right)^{1-i}\,
\sigma^{-k}\,(1-x_3)^m\, F_n[z_{\rm acn}]\,,
\nnnl
&&\nonumber\\
z_{\rm acn}&=&x_3^2+(1-x_3)\,\sigma-x_3\,(1-x_3)\,a\,,
\nonumber\\
a &=& x_1\,x_2\,s+x_1\,(t-1) +x_2\,(u-1)\,,
\label{Aikmn}
\end{eqnarray}
where $F_n[z]$ is defined in (\ref{eq:F_n}).
The integrals $A(i,k,m,n)$ are convergent at $w=0$
in the case
\begin{eqnarray*}
&&
i=1,\quad k=0,\quad m\ge 1,\quad n\ge 4\,,
\\ 
&&
i=1,\quad k\ge 1,\quad m\ge 1,\quad n\ge 3\,,
\\
&&
i=2,\quad k\ge 0,\quad m\ge 0,\quad n\ge3\,.
\end{eqnarray*}
 To single out the divergent part in the remaining integrals,
we invoke recursion relations in the 
following manner.
We perform a  partial integration in $x_3$, and use 
$$
\int dx_3\,(1-x_3)^m=-\frac{(1-x_3)^{m+1}}{m+1}\,.
$$
Then we express $(1-x_3)\,\sigma$ through $z_{\rm acn}$ and obtain
\begin{eqnarray}
&&(m+3-n+\omega)A(i,k,m,n)=
\nnnl
&&{\mbox {Div}}(i,k,n)
-n\left[A(i,k,m,n+1)- (1+a)\,A(i,k,m+2,n+1)\right]\,,
\label{recAcnode}
\end{eqnarray}
where
\begin{eqnarray}
{\mbox {Div}}(i,k,n)&=&4^{3-n-k+\omega}\,
\frac{\,C^2(w)\,\Gamma(3/2)}{\Gamma(3/2+\omega)}\,
\frac{\Gamma(n-2-\omega)}{\Gamma(n)}\times
\nnnl
&& B(5/2-i+\omega,n+k-4+i-2\omega)\, , \nnnl 
B(x,y)&=&\frac{\Gamma(x)\Gamma(y)}{\Gamma(x+y)}\,.
\end{eqnarray}
The divergences in $Div(i,k,n)$ can be worked out straightforwardly.

The integral $A(1,0,0,3)$ must be considered separately. We write 
\begin{eqnarray}\label{eq:a13}
A(1,0,0,3) &=& D(0,3)+\int_4^\infty [d\sigma] 
\int\limits_0^1\! d^3x\{F_3[z_{\rm acn}]-F_3[y]\}\,,\nnnl
y &=& x_3^2+(1-x_3)\,\sigma. \label{A1003}
\end{eqnarray}
The divergent quantity $D(0,3)$ is worked out in Appendix C.1 of 
Ref.~\cite{GS}, whereas the integral on the right-hand side  is
 convergent at $d=4$.

This concludes our discussion of the acnode integral (\ref{eq:acnode}).

\section{The quantities $\Delta_A$ and $\Delta_B$}\label{app:deltaab}
Here we display the expressions for the quantities $\Delta_{A(B)}$
in Eqs.~(\ref{UA}) and (\ref{UB}).

\begin{eqnarray}
&&
\Delta_{A} (s,t,u) = 
\frac{1}{(4\,\pi\,F_\pi)^4 }
\left\{ \left( \frac{689}{162}-\frac{4\,\pi^2}{9} \right)
       +\frac{15043}{64800}\,s \right\}
\nonumber\\
&&
+\frac{1}{(4\,\pi\,F_\pi)^4}\,\frac{1}{288}\, 
\left\{   F_A^{\rm acn}(  s,  t,  u)\,
       +\,F_A^{\rm ver}(  s)
       +\,F_A^{\rm box}(  s,  t,  u)\right\},
\label{delA}\\
&&\nonumber\\
&&
\Delta_{B} (s,t,u) = 
\frac{1}{(4\,\pi\,F_\pi)^4 }
\left\{ \frac{8329}{43200}
+\left(\frac{2987}{1350}-\frac{2\,\pi^2}{9}\right)\frac{1}{s}
\right\}
\nonumber\\
&&
+\frac{1}{(4\,\pi\,F_\pi)^4}\,\frac{1}{288} 
     \left\{F_B^{\rm acn}(  s,  t,  u)\,
         +\,F_B^{\rm ver}(  s)
       \,+\,F_B^{\rm box}(  s,  t,  u)\right\}\, ,
\label{delB}
\end{eqnarray}
where 
\begin{eqnarray}
F_{I}^{\rm acn} &=& \int\limits_4^\infty\! d\sigma \beta\!
                    \int\limits_0^1\! d^3x
\left\{
\left[ \frac{P_{I;\,{\rm acn}}^{(0)}}{y}\,
    +\,\frac{P_{I;\,{\rm acn}}^{(1)}}{\sigma} 
\right]\,\frac{1}{z_{\rm acn}} 
    +\,\frac{P_{I;\,{\rm acn}}^{(2)}}{z^2_{\rm acn}}
\right\},
\nnnl
&&\nonumber\\
&&\nonumber\\
F_{I}^{\rm ver} &=& 
\int\limits_4^\infty\! \frac{d\sigma\,\beta}{\sigma}\!
\int\limits_0^1\! d^2x
\cdot\frac{P_{I;\,{\rm ver}}}{z_{\rm ver}}\, ,
\nnnl
&&\nonumber\\
&&\nonumber\\
F_{I}^{\rm box} &=& 
\!\!
\int\limits_4^\infty\! \frac{d\sigma\,\beta}{\sigma}
\int\limits_0^1\! d^3x
\sum\limits_{n=1}^2
\left\{
 P_{I;\, \rm{box_+}}^{(n)} D^{(n)}_{\rm box_+} 
+P_{I;\, \rm{box_-}}^{(n)} D^{(n)}_{\rm box_-}
\right\}\, ; \, I=A,B\, ,
\nnnl
\eea
and
\bea
D^{(n)}_{\rm box_\pm} &=&
\frac{1}{zn_{\rm box;\,t}} \pm \frac{1}{z^n_{\rm box;\,u}}\, .
\nonumber\end{eqnarray}
Here $P_{I}$  are polynomials in 
 $s, \nu=t-u$ and in $x_i$. Their explicit expressions are given in 
Appendix \ref{app:polynomials}.
 The quantity $z_{\rm acn}$ is displayed in Eq.~(\ref{Aikmn}), $y$ is given in
(\ref{eq:a13}), and
\bea
z_{\rm ver} &=& \sigma\,(1-x_3) +x_3^2\, y_2,
\hspace{1cm} y_2 = 1-s\,x_2\,(1-x_2),
\nonumber\\
z_{\rm box;\,t} &=& B_{\,\rm t}-A_{\,\rm t}\,x_1,\nnnl
A_{\,\rm t} &=& x_2\,x_3\,\left[s\,(1-x_2)\,x_3 + (1-t)\,(1-x_3)\right],
\nonumber\\
B_{\,\rm t}&=&A_{\,\rm t} + z_{\rm ver}\,,\nnnl
z_{\rm box;\,u}&=&z_{\rm box;\,t}|_{t\rightarrow u}\,.
\eea
The acnode integrals are easy to evaluate numerically
in the physical region for the reaction $\gamma\gamma\to\pi\pi$, because
 branch points occur at $t= 4,u=4$ only.

On the other hand, the vertex and box integrals contain branch 
 points at $s=4$.
 In order to evaluate these integrals at $s\ge 4$, we invoke dispersion
 relations in the manner described in \cite{GS}.

\section{The polynomials $P_A$ and $P_B$}
\label{app:polynomials}
Here, we display the polynomials $P_{A(B)}$ that occur in the expressions
$\Delta_{A(B)}$ in Appendix \ref{app:deltaab}. We use the abbreviations
\bea
x_{+} &=& x_1+x_2-2\, x_1\, x_2\,,\, x_{-} = x_1-x_2\, ,\nnnl
x_{123}&=&(1+x_3-2\,x_2\, x_3)(1-x_3+2\,x_1\,x_2\,x_3)\, .
\eea

\subsection{The polynomials $P_A$}
\begin{eqnarray*}
&&
P_{\rm A;\, acn}^{(0)} =
-\,192\,x_3\,(1-x_3)(s\,x_{+} -\nu\, x_{-}), 
\\
&&\\
&&
P_{\rm A;\, acn}^{(1)} =-\, 6\,s\,\nu\, x_{-}\,(1-x_3)
\\
&&
\times
\left[1 + 8\, x_3^3 + 3\, x_3^4 
      - 4\, x_{+}\, (x_3 + 3\, x_3^3 + 2\, x_3^4)
      + 2\, x^2_{-}\,(1 +2\, x_3^3 -3\, x_3^4)
\right]
\\
&&
- 6\,s^2\,(1-x_3)
\left[ 
x_{+}^2\, (1 + 2\, x_3 + 14\, x_3^3 + 7\, x_3^4) + 6\,  x_{-}^2\, x_3^4 
\right.
\\
&&
\left.
 -  x_{+}\,\left(1 + 2\, x_{-}^2 + 4\, x_3^3\, (2 + x_{-}^2) 
  + 3\, x_3^4\, (1 + 2 x_{-}^2)\right)
\right]
\\
&&
+ 6\,\nu^2\,x_{-}^2\,(1-x_3)\,
\left[ 
1 - 2\, x_3 + 2\, x_3^3 + 5\, x_3^4 - 12\, x_{+} x_3^4
\right]
\\
&&
+ 12\,s
\left[
- 2\, x_{+}^2\, (1 - x_3)^3\, (1 + 2 x_3 + 3 x_3^2)
\right.
\\
&& 
+  x_3\, (2 + 33\, x_3 - 19\, x_3^2 - 19\, x_3^3 + 15\, x_3^4) 
\\
&&
+  x_{+}\, (-4 + 2\, x_3 - 63\, x_3^2 + 25\, x_3^3 + 51\, x_3^4 - 35\, x_3^5) 
\\
&&
\left.
  +    2\,  x_{-}^2\, x_3\, (2 + 9\, x_3 - 7\, x_3^2 + 5\, x_3^3 - 3\, x_3^4)
\right]
\\
&&
+ 12\,\nu\,x_{-}\,(1-x_3)\,
\left[
4 - 2\, x_3 - 5\, x_3^2 - 4\, x_3^3 + 19\, x_3^4 
\right.
\\
&&
\left.
+ 2\, x_{+}\, (1 + 8\, x_3^3 - 21\, x_3^4)
\right]
\\
&&
- 48\,x_3
\left[ 4 + 4\, x_3 + 6\, x_3^2 - 19\, x_3^3 + 10\, x_3^4 
\right.
\\
&&
\left.
+  x_{+}\, (2 - 3\, x_3 - 19\, x_3^2 + 41\, x_3^3 - 21\, x_3^4) 
\right],
\\
&&\\
&&
P_{\rm A;\, acn}^{(2)} =
2\,s\,\nu\, x_{-}\,(1-x_3)^2\,
\left[
-  11 - 8\, x_3 - 8\, x_3^2 - 24\, x_3^3 - 9\, x_3^4 
\right.
\\
&&
\left.
+  12\, x_{+}\, (4 + x_3 + 3\, x_3^3 + 2\, x_3^4) 
-  2\, x_{-}^2 \,(11 + 8\, x_3 + 8\, x_3^2 + 6\, x_3^3 - 9\, x_3^4) 
\right]
\\
&&
- 2\,s^2\,(1-x_3)^2
\left[
x_{+}^2 (35 - 18\, x_3 + 34\, x_3^3 + 21\, x_3^4) + 18\, x_{-}^2\, x_3^4  
\right.
\\
&&
\left.
- x_{+}\, \left(11 + 16\, x_3^3 + 9\, x_3^4 
            +2\, x_{-}^2\, (11- 2\, x_3^3 +9\, x_3^4)\right) 
\right]
\\
&&+2\, \nu^2\,x_{-}^2\,(1-x_3)^2\,
\left[
-13 + 18\, x_3 - 2\, x_3^3 + (15 - 36\, x_{+})\, x_3^4
\right]
\\
&&+ 4\,s\,(1-x_3)\,
\left[
 x_3\, (22 + 11\, x_3 + 83\, x_3^2 - 101\, x_3^3 + 45\, x_3^4) 
\right.
\\
&&
- 2\, x_{+}^2\, (1 - x_3)^2 \,
\left(11(1 + x_3 + x_3^2) - 9\, x_3^3\right) 
\\
&&
+  x_{+}\, (60 - 178\, x_3 + 43\, x_3^2 - 173\, x_3^3 + 233\, x_3^4 
- 105\, x_3^5) 
\\
&&
\left.
+  2\, x_{-}^2\, (2 - x_3)\, x_3\, 
\left(11(1 +  x_3 + x_3^2) + 9\, x_3^3\right) 
\right]
\\
&&+ 4\,\nu\,x_{-}\,(1-x_3)^2\,
\left[
-60 + 74\, x_3 + 9\, x_3^2 - 56\, x_3^3 + 57\, x_3^4 
\right.
\\
&&
\left.
+  x_{+}\, (22 + 104\, x_3^3 - 126\, x_3^4)
\right]
\\
&&
+ 16\,x_3\,(1-x_3)
\left[
60 + 18\, x_3 - 108 \,x_3^2 + 93\, x_3^3 - 30\, x_3^4 
\right.
\\
&&
\left.
+ x_{+}\, (-22 - 83 \,x_3 + 205\, x_3^2 - 187\, x_3^3 + 63\, x_3^4)
\right],
\\
&&\\
&&
P_{\rm A;\, ver} =
128\,s\,x_2^2\,(1-2\,x_2)\,x_3^4
\left[3 - 2\, x_2\,(2 - 15\, x_3)\, x_3 
- 18\, x_3^2 - 15\, x_2^2\, x_3^2\right]
\\
&&- 32\,s^2\,
x_2^2\, (1 - 2\, x_2)\,(30 - 42\, x_2 + 7\, x_2^2)\, x_3^6 \, ,
\\
&&\\
&&
P^{(1)}_{\rm A;\,box_+}=
- 96\,s\,x_2\,x_3^2
\left[
-6 + 9\, x_3 + 20\, x_2\, x_3 - 22\, x_1\, x_2\, x_3 
\right.
\\
&&
- 2\, x_3^2 - 40\, x_2\, x_3^2 + 6\, x_1\, x_2\, x_3^2 - 4\, x_2^2\, x_3^2 
+ 52\, x_1\, x_2^2\, x_3^2 
- 8\, x_1^2\, x_2^2\, x_3^2 
\\
&&
- 15\, x_3^3 + 58\, x_2\, x_3^3 + 14\, x_1\, x_2\, x_3^3 
-12\, x_2^2\, x_3^3 - 72\, x_1\, x_2^2\, x_3^3 - 2\, x_1^2\, x_2^2\, x_3^3 
\\
&&
+ 8\, x_1\, x_2^3\, x_3^3 
+ 4\, x_1^2\, x_2^3\, x_3^3 + 20\, x_3^4 
- 48\, x_2\, x_3^4 - 36\, x_1\, x_2\, x_3^4 + 20\, x_2^2\, x_3^4 
\\
&&
\left.
+76\, x_1\, x_2^2\, x_3^4 
+ 10\, x_1^2\, x_2^2\, x_3^4 
- 24\, x_1\, x_2^3\, x_3^4  - 24\, x_1^2\, x_2^3\, x_3^4 
+ 24\, x_1^2\, x_2^4\, x_3^4
\right]
\\
&&
- 16\,s^2\,x_2^2\,x_3^4 
\left[
12 + 12\, x_1 - 12\, x_2 - 24\, x_1\, x_2 + 12\, x_1^2\, x_2 + 30\, x_1\, x_3
\right.
\\
&&
 -  24\, x_2\, x_3 - 90\, x_1\, x_2\, x_3 + 27\, x_1^2\, x_2\, x_3 
+ 30\, x_2^2\, x_3 +     66\, x_1\, x_2^2\, x_3 
\\
&&    
    - 24\, x_1^2\, x_2^2\, x_3 - 10\, x_1^3\, x_2^2\, x_3 - 
    24\, x_3^2 + 6\, x_1\, x_3^2 + 66\, x_2\, x_3^2 - 18\, x_1\, x_2\, x_3^2 
\\
&&    
+  45\, x_1^2\, x_2\, x_3^2 - 48\, x_2^2\, x_3^2 + 48\, x_1\, x_2^2\, x_3^2 - 
    120\, x_1^2\, x_2^2\, x_3^2 - 28\, x_1^3\, x_2^2\, x_3^2\\  
&&    
\left.
-24\, x_1\, x_2^3\, x_3^2 + 36\, x_1^2\, x_2^3\, x_3^2 
+ 56\, x_1^3\, x_2^3\, x_3^2
\right]
\\
&&
+ 48\,\nu^2\,x_2^3\,x_3^4 
\left[
-4 + 8\, x_1 - 8\, x_1^2 + 2\, x_1\, x_3 + 7\, x_1^2\, x_3 + 10\, x_2\, x_3 
\right.
\\
&&
-    30\, x_1\, x_2\, x_3 + 30\, x_1^2\, x_2\, x_3 - 10\, x_1^3\, x_2\, x_3 
+ 6\, x_3^2 -  14\, x_1\, x_3^2 + 3\, x_1^2\, x_3^2 
\\
&&
- 12\, x_2\, x_3^2 + 12\, x_1\, x_2\, x_3^2 + 12\, x_1^2\, x_2\, x_3^2 
- 12\, x_1^3\, x_2\, x_3^2 + 24\, x_1\, x_2^2\, x_3^2 
\\
&&
\left.
-  48\, x_1^2\, x_2^2\, x_3^2 + 24\, x_1^3\, x_2^2\, x_3^2
\right]
\\
&&
+ 384\,x_2\,x_3^2
\left[
-6 + 9\, x_3 + 19\, x_2\, x_3 - 19\, x_1\, x_2\, x_3 + 7\, x_3^2 
- 50\, x_2\, x_3^2
\right. 
\\
&&
-  2\, x_1\, x_2\, x_3^2 + 52\, x_1\, x_2^2\, x_3^2 - 15\, x_3^3 
+ 44\, x_2\, x_3^3 +     16\, x_1\, x_2\, x_3^3 \\
&&    
    - 60\, x_1\, x_2^2\, x_3^3 + 8\, x_3^4 - 16\, x_2\, x_3^4 - 
    16\, x_1\, x_2\, x_3^4 + 32\, x_1\, x_2^2\, x_3^4
\left.\right] ,
\\
&&\\
&&
P^{(2)}_{\rm A;\,box_+} =
-24\,s\,\nu^2\,(1-x_1)^2\,x_2^3\,
(1 + x_2 + x_1\, x_2 - 2\, x_1\, x_2^2)\,x_3^6\,x_{123}
\\
&&
+ 96\,s\,x_2\,x_3^4\,x_{123}
\left[
-7\, x_2 - 7\, x_1\, x_2 + 14\, x_1\, x_2^2 
+ 3\, x_3 + 6\, x_2\, x_3 
\right.
\\
&&
\left.
+ 6\, x_1\, x_2\, x_3 - 12\, x_1\, x_2^2\, x_3 - 3\, x_3^2 - 
        2\, x_2\, x_3^2 - 2\, x_1\, x_2\, x_3^2 + 4\, x_1\, x_2^2\, x_3^2
\right]
\\
&&
- 48\,s^2\,x_2^2\, x_3^4\,x_{123}
\left[
  -2 - 2\, x_1 - x_2 + 6\, x_1\, x_2 - x_1^2\, x_2 + 
 3\, x_3 + 3\, x_1\, x_3 
\right.
\\
&&      
        - 6\, x_1\, x_2\, x_3 - 4\, x_3^2 - 4\, x_1\, x_3^2 + 
        2\, x_2\, x_3^2 + 8\, x_1\, x_2\, x_3^2 \\
&&
\left.      
        + 2\, x_1^2\, x_2\, x_3^2 - 
        4\, x_1\, x_2^2\, x_3^2 - 4\, x_1^2\, x_2^2\, x_3^2 + 
        4\, x_1^2\, x_2^3\, x_3^2
\right]
\\
&&
- 24\,s^3\,x_2^3\,
(1 + x_1 - 2\, x_1\, x_2)\, (1 + x_1 - x_2 - x_1^2\, x_2)\,
x_3^6\,x_{123} 
\\
&&
- 48\,\nu^2\,(1-x_1)^2\, x_2^3\,(1-x_3)\,\,x_3^4\,
(1+4\,x_3)\,x_{123}
+2\, x_2\, (1 - x_3)^2\, x_3^4\,x_{123}\,,
\\
&&\\
&&
P^{(1)}_{\rm A;\,box_-} =
-\,32\,s\,\nu\,x_2^2\, x_3^4
\left[
-6 + 6\, x_1 + 12\, x_1\, x_2 - 18\, x_1^2\, x_2
\right.
\\
&& 
 - 15\, x_1\, x_3 + 12\, x_2\, x_3 + 
33\, x_1^2\, x_2\, x_3 - 30\, x_1\, x_2^2\, x_3 + 21\, x_1^2\, x_2^2\, x_3 
\\
&&    
- 10\, x_1^3\, x_2^2\, x_3 + 12\, x_3^2 + 3\, x_1\, x_3^2 - 24\, x_2\, x_3^2 - 
    30\, x_1\, x_2\, x_3^2 
\\
&&    
\left.
    + 18\, x_1^2\, x_2\, x_3^2 + 6\, x_2^2\, x_3^2 + 
54\, x_1\, x_2^2\, x_3^2 - 48\, x_1^2\, x_2^2\, x_3^2 
+ 4\, x_1^3\, x_2^2\, x_3^2
\right]
\\
&&
- 96\,\nu\,x_2^2\,x_3^3
\left[
-9+ 11\, x_1 + 12\, x_3 - 14\, x_1\, x_3 + 24\, x_2\, x_3 
\right.
\\
&&
- 56\, x_1\, x_2\, x_3 + 28\, x_1^2\, x_2\, x_3 + 15\, x_3^2 
- 27\, x_1\, x_3^2 - 58\, x_2\, x_3^2 
\\
&&    
    +  84\, x_1\, x_2\, x_3^2 
+ 8\, x_1^2\, x_2\, x_3^2 + 60\, x_1\, x_2^2\, x_3^2 - 
    80\, x_1^2\, x_2^2\, x_3^2 
\\
&&    
    - 16\, x_3^3 + 28\, x_1\, x_3^3 + 32\, x_2\, x_3^3 - 
20\, x_1\, x_2\, x_3^3 - 42\, x_1^2\, x_2\, x_3^3 
\\
&&
\left.
- 64\, x_1\, x_2^2\, x_3^3 + 84\, x_1^2\, x_2^2\, x_3^3
\right],
\\
&&\\
&&
P^{(2)}_{\rm A;\, box_-} =
48\,s\,\nu\,(1-x_1)\,x_2^2\,x_3^4\,x_{123}
\\
&&
\times
\left[
-2 +   3\, x_3 + 3\, x_2\, x_3 + 3\, x_1\, x_2\, x_3 
- 6\, x_1\, x_2^2\, x_3 - 4\, x_3^2 
\right.
\\
&&
\left.
-  2\, x_2\, x_3^2 - 2\, x_1\, x_2\, x_3^2 
+ 4\, x_1\, x_2^2\, x_3^2
\right]      
\\
&&
+ 24\,s^2\,\nu\,(1-x_1^2)\,
 x_2^3\, (2 - x_2 - x_1\, x_2)\, x_3^6\,x_{123} 
\\
&&
+\nu\, (1 - x_1)\, x_2^2\, (1 - x_3)\, x_3^4\, 
(1 -  2\, x_3)\, x_{123}
\\
&&
+ 24\,\nu^3\,
(1 - x_1)^3\, x_2^4\, x_3^6\,x_{123}\,.
\end{eqnarray*}

\subsection{The polynomials $P_B$}
\begin{eqnarray*}
&&
P_{\rm B;\, acn}^{(0)} =
96\,\,x_3\,(1-x_3)\,
\left[\frac{\nu}{s}\, x_{-} - x_{+} \right], 
\\
&&\\
&&
P_{\rm B;\, acn}^{(1)} =
\frac{6\,\nu}{s}\,x_{-}\,
\left[  
4-2\, x_3+21\, x_3^2-19\, x_3^3+3\, x_3^4+5\, x_3^5
\right]
\\
&&
- \frac{3\,\nu^2}{s}\,\,x_{-}^2\,(1-x_3)\,[1+2\, x_3+2\, x_3^3 + x_3^4]
\\
&&
- \frac{24}{s}\,x_3\,(2-x_3^2)\,[2 -8\, x_3 + 5\, x_3^2]
+ 3\,s\,(1 - x_3)\,
\\
&&
\times\left[
 x_{+}^2\, (1-x_3)^3\, (1+x_3) -x_{+}\, (1+8 \,x_3^3 +3\, x_3^4)
+ 6\,x_{-}^2\, x_3^4
\right]
\\
&&
+ 6\,\left[
x_{+}\, (1-x_3)^2\, (-4 - 2\, x_3 - 3\, x_3^2 + 13\, x_3^3) 
\right.
\\
&&
\left.
 +  x_3\, (-2 - 33\, x_3 + 19\, x_3^2 + 19\, x_3^3 - 15\, x_3^4)
\right],
\\
&&\\
&&
P_{\rm B;\, acn}^{(2)} =
-\,\frac{2\,\nu}{s}\,x_{-}\,(1-x_3)\,
\\
&&
\times
\left[
-34 + 152\, x_3 - 43\, x_3^2 + 21\, x_3^3 - 27\, x_3^4 + 15\, x_3^5
\right]
\\
&&
-\frac{\nu^2}{s}\,x_{-}^2\,(1-x_3)^2
\left[35 - 18\, x_3 - 2\, x_3^3 + 3\, x_3^4 \right]
\\
&&
+ \frac{8}{s}\,x_3\,(1-x_3)
\left[34 - 67\, x_3 + 95\, x_3^2 - 56\, x_3^3 + 15\, x_3^4\right]
\\
&&
- s\,(1-x_3)^2
\left[
 x_{+}^2\, (1-x_3)^2\, (13 + 8\, x_3+3\, x_3^2)
\right.
\\
&&
\left.
+ x_{+}\, (11+16\, x_3^3 +9\, x_3^4) - 18\, x_{-}^2\, x_3^4
\right] 
\\
&&+ \nu\,x_{-}\,(1-x_3)^2\,
\left[
  11+16\, x_3^3 +9\, x_3^4 
+ 12\, x_{+}\, (4 - 3\, x_3 - x_3^4)
\right]
\\
&&
+ 2\,(1-x_3)
\left[
-x_3 \,(22 + 11\, x_3 + 83\, x_3^2 - 101\, x_3^3 + 45\, x_3^4) 
\right.
\\
&&
\left.
+  x_{+}\, (34 - 108\, x_3 + 65\, x_3^2 + 73\, x_3^3 - 103\, x_3^4 + 
            39\, x_3^5)
\right],
\\
&&\\
&&
P_{\rm B;\, ver} =
16\,s\,x_2^2\,(1-2\,x_2)\,x_3^5
\left[ 33\, x_2^2\, x_3 - 6\,(4 - 5\, x_3) - 2\, x_2\,(4 + 9\, x_3)\right]\\
&&
- 1536\,x_2^2(1-2\,x_2)\,(1-x_3)\,(1-2\,x_3)\,x_3^4\,,
\\
&&\\
&&
P^{(1)}_{\rm B;\, box_+} =
- \frac{24\,\nu^2}{s}\,x_2^3\,(1-x_3)\,x_3^4
\left[
4 + 8\, x_1 - 10\, x_3 - 2\, x_1\, x_3 - 3\, x_1^2\, x_3
\right]
\\
&&
-\frac{ 1152}{s}\,x_2\,(1-x_3)^2\,x_3^2\,(1-2\,x_3+2\,x_3^2)
\\
&&
- 24\,s\,x_2^2\,x_3^4
\left[
4 + 12\, x_1 - 4\, x_2 - 12\, x_3 - 14\, x_1\, x_3 + 10\, x_2\, x_3
\right.
\\
&& 
 - 14\, x_1\, x_2\, x_3 + 
17\, x_1^2\, x_2\, x_3 - 14\, x_1^2\, x_2^2\, x_3 
+ 8\, x_3^2 - 2\, x_1\, x_3^2 
     -  6\, x_2\, x_3^2    \\
&&
\left.    
    + 14\, x_1\, x_2\, x_3^2 - 3\, x_1^2\, x_2\, x_3^2 + 
 4\, x_1\, x_2^2\, x_3^2 - 12\, x_1^2\, x_2^2\, x_3^2 
+ 12\, x_1^2\, x_2^3\, x_3^2
\right]
\\
&&
- 48\,x_2\,(1-x_3)\,x_3^2 \left[
-6 + 19\, x_3 + 16\, x_1\, x_2\, x_3 - 27\, x_3^2 - 2\, x_2\, x_3^2 
\right.
\\
&&
\left.
- 18\, x_1\, x_2\, x_3^2 - 28\, x_1\, x_2^2\, x_3^2 + 20\, x_3^3 - 
    32\, x_1\, x_2\, x_3^3 + 64\, x_1\, x_2^2\, x_3^3
\right],
\\
&&\\
&&
P^{(2)}_{\rm B;\, box_+} =
-\,\frac{24\,\nu^2}{s}\,(1-x_1)^2\,
x_2^3\,(1 - x_3)^3\, x_3^4\, (1 + 4\, x_3)
\\
&&
+\,\frac{1}{s}\,x_2\,(1-x_3)^4\,x_3^4
\\
&&
- 24\,s\,x_2^2\,(1-x_3)^2\,x_3^4
\left[
-2 - 2\, x_1 - x_2 + 6\, x_1\, x_2 - x_1^2\, x_2 
\right.
\\
&&
+ 3\, x_3 + 3\, x_1\, x_3 - 
    6\, x_1\, x_2\, x_3 - 4\, x_3^2 - 4\, x_1\, x_3^2 + 2\, x_2\, x_3^2 \\
&&
\left.    
    +  8\, x_1\, x_2\, x_3^2 + 2\, x_1^2\, x_2\, x_3^2 
- 4\, x_1\, x_2^2\, x_3^2 - 
    4\, x_1^2\, x_2^2\, x_3^2 + 4\, x_1^2\, x_2^3\, x_3^2
\right]
\\
&&
- 12\,s^2\,x_2^3\,(1 + x_1 - 2\ x_1\ x_2)\,(1 + x_1 - x_2 - x_1^2\, x_2)\,
(1 - x_3)^2\, x_3^6
\\
&&
+ 12\,\nu^2\,(1 - x_1)^2\,x_2^3\,(-1 - x_2 - x_1\, x_2 + 
        2\, x_1\, x_2^2)\,(1 - x_3)^2\, x_3^6
\\
&&+ 48\,x_2\,(1-x_3)^2\,x_3^4 
\left[
-7\, x_2 - 7\, x_1\, x_2 + 14\, x_1\, x_2^2 + 3\, x_3 
\right.
\\
&&
 + 6\, x_2\, x_3 + 
 6\, x_1\, x_2\, x_3 - 12\, x_1\, x_2^2\, x_3 - 3\, x_3^2 - 2\, x_2\, x_3^2 
\\
&&
\left.
    -  2\, x_1\, x_2\, x_3^2 + 4\, x_1\, x_2^2\, x_3^2
\right],   
\\
&&\\
&&
P^{(1)}_{\rm B;\, box_-} =
\frac{48\,\nu}{s}\,x_2^2\,(1-x_3)^2\,x_3^3\,
 ( 5 + 11\, x_1 - 28\, x_3 - 4\, x_1\, x_3)
\\
&&
+ 48\,\nu\,x_2^2\,x_3^4
\left[
2 + 2\, x_1 + 8\, x_1\, x_2 - 6\, x_3 - 3\, x_1\, x_3 - 2\, x_2\, x_3 
\right.
\\
&&
  -  4\, x_1\, x_2\, x_3 + 3\, x_1^2\, x_2\, x_3 - 12\, x_1\, x_2^2\, x_3 
  -  3\, x_1^2\, x_2^2\, x_3 + 4\, x_3^2 
\\
&&      
\left.
      + x_1\, x_3^2 + 2\, x_2\, x_3^2 - 
      10\, x_1\, x_2\, x_3^2 + 18\, x_1\, x_2^2\, x_3^2
\right],
\\
&&\\
&&
P^{(2)}_{\rm B;\, box_-} =
 \frac{144\,\nu}{s}\,
(1 - x_1)\, x_2^2\, (1 - x_3)^3\, x_3^4\, (1\, - 2\, x_3)\\
&&
+ \frac{12\,\nu^3}{s}\,(1-x_1)^3\,x_2^4\,(1-x_3)^2\,x_3^6
\\
&&- 12\,s\,\nu\,(1-x_1^2)\,x_2^3\,
(-2 + x_2 + x_1\, x_2)\, (1 - x_3)^2\, x_3^6
\\
&&
+ 24\, \nu\,(1-x_1)\,x_2^2\,(1 - x_3)^2\, x_3^4\,
\left[
-2 + 3\, x_3 + 3\, x_2\, x_3 + 3\, x_1\, x_2\, x_3 
\right.
\\
&&
\left.
- 6\, x_1\, x_2^2\, x_3 - 4\, x_3^2 
- 2\, x_2\, x_3^2 -2\, x_1\, x_2\, x_3^2 + 4\, x_1\, x_2^2\, x_3^2
\right].
\end{eqnarray*}


\begin{thebibliography}{999}

\bibitem{GLann}
  J.~Gasser and H.~Leutwyler,
  Annals Phys.\  {\bf 158} (1984) 142.

\bibitem{GLnpb}
  J.~Gasser and H.~Leutwyler,
  Nucl.\ Phys.\ B {\bf 250} (1985) 465.

\bibitem{BijnensCornet}
  J.~Bijnens and F.~Cornet,
  Nucl.\ Phys.\ B {\bf 296} (1988) 557.

\bibitem{Donoghuepipi}
  J.~F.~Donoghue, B.~R.~Holstein and Y.~C.~Lin,
  Phys.\ Rev.\ D {\bf 37} (1988) 2423 .

\bibitem{BGS}
  S.~Bellucci, J.~Gasser and M.~E.~Sainio,
  Nucl.\ Phys.\ B {\bf 423} (1994) 80  \newline
  [Erratum-ibid.\ B {\bf 431} (1994) 413]
  [arXiv:hep-ph/9401206].

\bibitem{Burgi} 
  U.~Burgi,
  Nucl.\ Phys.\ B {\bf 479} (1996) 392 
  [arXiv:hep-ph/9602429];\\
  U.~Burgi,
  Phys.\ Lett.\ B {\bf 377} (1996)  147
  [arXiv:hep-ph/9602421].

\bibitem{SchererFearing}
  H.~W.~Fearing and S.~Scherer,
  Phys.\ Rev.\ D {\bf 53} (1996) 315 \newline
  [arXiv:hep-ph/9408346].

\bibitem{BCE1}
  J.~Bijnens, G.~Colangelo and G.~Ecker,
  JHEP {\bf 9902} (1999) 020  \newline
  [arXiv:hep-ph/9902437].

\bibitem{BCE2}
  J.~Bijnens, G.~Colangelo and G.~Ecker,
  Annals Phys.\  {\bf 280} (2000) 100 \newline
  [arXiv:hep-ph/9907333].

\bibitem{GS}
  J.~Gasser and M.~E.~Sainio,
  Eur.\ Phys.\ J.\ C {\bf 6} (1999) 297 \newline
  [arXiv:hep-ph/9803251].

\bibitem{Guiasu Radescu}
 I.~Guiasu and E.~E.~Radescu,
  Annals Phys.\  {\bf 120} (1979) 145; ibid.\, {\bf 122} (1979) 436.

\bibitem{Filkovquadrupole}
  L.~V.~Fil'kov and V.~L.~Kashevarov,
  arXiv:nucl-th/0505058.

\bibitem{MAMI}
  J.~Ahrens {\it et al.},
  Eur.\ Phys.\ J.\ A {\bf 23} (2005) 113
  [arXiv:nucl-ex/0407011].


\bibitem{COMPASS} 
The COMPASS Collaboration, 
CERN Proposal CERN/SPSLC 96-14, SPSC/P 297, March 1,
 1996, and 
 Addendum 1, CERN/SPSLC 96-30, SPSC/P 297, March 20, 1996\newline
[http://wwwcompass.cern.ch/compass/proposal/welcome.html].



\bibitem{COMPASS1}
  M.~Moinester  [for the COMPASS Collaboration],
Workshop {\it Symmetries and Spin}-Praha-SPIN-2002, Advanced Study Institute,
INTAS monitoring Conference, Prague, 14-24 July, 2002, 
  Czech.\ J.\ Phys.\  {\bf 53} (2003) B169
  [arXiv:hep-ex/0301024].


\bibitem{pioncharged}
J.~Gasser, M.A.~Ivanov, M.E.~Sainio, work in progress. 

\bibitem{CGLpipi}
  G.~Colangelo, J.~Gasser and H.~Leutwyler,
  Nucl.\ Phys.\ B {\bf 603} (2001) 125 \newline
  [arXiv:hep-ph/0103088].

\bibitem{BijnensTalavera}
  J.~Bijnens and P.~Talavera,
  Nucl.\ Phys.\ B {\bf 489} (1997) 387  \newline
  [arXiv:hep-ph/9610269].


\bibitem{KnechtMoussallamStern}
  M.~Knecht, B.~Moussallam and J.~Stern,
  Nucl.\ Phys.\ B {\bf 429} (1994) 125 \newline
  [arXiv:hep-ph/9402318].

\bibitem{Holsteinfpi}
  B.~R.~Holstein,
  Phys.\ Lett.\ B {\bf 244} (1990) 83.

\bibitem{Descotesfpi}
  S.~Descotes-Genon and B.~Moussallam,
  arXiv:hep-ph/0505077.


\bibitem{Vermaseren}
  J.~A.~M.~Vermaseren,
  arXiv:math-ph/0010025.

\bibitem{Burgithesis}
U.~Burgi,
Ph.~D. thesis, University of Bern, 1996.

\bibitem{Filkovdipole}
  L.~V.~Fil'kov and V.~L.~Kashevarov,
  Eur.\ Phys.\ J.\ A {\bf 5} (1999) 285 \newline 
  [arXiv:nucl-th/9810074].

\bibitem{Kaloshindipole}
  A.~E.~Kaloshin, V.~M.~Persikov and V.~V.~Serebryakov,
  Phys.\ Atom.\ Nucl.\  {\bf 57} (1994) 2207
  [Yad.\ Fiz.\  {\bf 57N12} (1994) 2298]
  [arXiv:hep-ph/9402220].

\bibitem{Bellucci}
  R. Baldini, S.~Bellucci, 
Proc. Workshop on Chiral Dynamics: Theory and Experiments, 
 Cambridge, MA, USA, 25-29 July, 1994, Lecture Notes in Physics 452 
(Springer Berlin, A.M. Bernstein, B.R. Holstein (eds.)), p.177;\\
 S. Bellucci, arXiv:hep-ph/9508282;\newline
  D.~Babusci, S.~Bellucci, G.~Giordano and G.~Matone,\newline
  Phys.\ Lett.\ B {\bf 314} (1993) 112; \newline
  D.~Babusci, S.~Bellucci, G.~Giordano, G.~Matone, A.~M.~Sandorfi and 
  M.~A.~Moinester,
  Phys.\ Lett.\ B {\bf 277} (1992) 158.

\bibitem{LECs_asymptotic}
  G.~Ecker, J.~Gasser, H.~Leutwyler, A.~Pich and E.~de Rafael,\newline
  Phys.\ Lett.\ B {\bf 223} (1989) 425;\newline
  G.~Ecker, J.~Gasser, A.~Pich and E.~de Rafael,
  Nucl.\ Phys.\ B {\bf 321} (1989) 311;
  M.~Knecht and A.~Nyffeler,
  Eur.\ Phys.\ J.\ C {\bf 21} (2001) 659 \newline
  [arXiv:hep-ph/0106034];\newline
  V.~Cirigliano, G.~Ecker, M.~Eidemuller, R.~Kaiser, 
A.~Pich and J.~Portoles,
  JHEP {\bf 0504} (2005) 006
  [arXiv:hep-ph/0503108].

\end{thebibliography}
\end{document}